\documentclass[twocolumn]{jpsj2}

\def\runauthor{
Akira \textsc{Oguri}, Yoshihide \textsc{Tanaka}, 
and A. C. \textsc{Hewson}
}

\title{
Quantum phase transition in a 
minimal model for 
the Kondo effect in a Josephson junction 
}

\author{
Akira \textsc{Oguri}, Yoshihide \textsc{Tanaka}, 
and A. C. \textsc{Hewson}$^1$
}%
\inst{%
Department of Material Science, 
Osaka City University, Sumiyoshi-ku, Osaka 558-8585, Japan \\
$^1$Department of Mathematics, Imperial College, 
180 Queen's Gate, London SW7 2BZ, UK
}%

\recdate{June 12, 2004}

\abst{
We propose a minimal model for 
the Josephson current through a quantum dot in a Kondo regime.
We start with the model that consists of an Anderson impurity connected to 
two superconducting (SC) leads 
with the gaps $\Delta_{\alpha}=|\Delta_{\alpha}|
\,\text{e}^{\text{i} \theta_{\alpha}}$, 
where $\alpha = L,\,R$ for the lead at left and right.
We show that, 
when one of the SC gaps is much larger 
than the others $|\Delta_L| \gg |\Delta_R|$,
the starting model can be mapped exactly onto 
the single-channel model, which consists of the right lead of $\Delta_R$ 
and the Anderson impurity with an extra onsite SC 
gap of  $\Delta_d \equiv \Gamma_L\,\text{e}^{\text{i} \theta_L}$. 
Here $\theta_L$ and $\Gamma_L$ are defined with 
respect to the starting model,  
and $\Gamma_L$ is the level width 
due to the coupling with the left lead. 
Based on this simplified model,
we study the ground-state properties for 
the asymmetric gap,  $|\Delta_L| \gg |\Delta_R|$, using  
the numerical renormalization group (NRG) method.  
The results show that 
the phase difference of the SC gaps $\phi \equiv \theta_R -\theta_L$,
which induces the Josephson current, 
disturbs the screening of the local moment 
to destabilize the singlet ground state typical of the Kondo system. 
It can also drive the quantum phase transition 
to a magnetic doublet ground state, and at the critical point 
the Josephson current shows a discontinuous change. 
The asymmetry of the two SC gaps causes 
a re-entrant magnetic phase, 
in which the in-gap bound state lies close to the Fermi level.
}

\kword{
Kondo effect, Josephson effect, 
quantum phase transition,  Anderson model, 
numerical renormalization group, quantum dot 
}

\begin{document}
\sloppy
\maketitle

\section{Introduction}
\label{sec:intro}

The Kondo effect in superconducting (SC) materials 
was originally studied for dilute magnetic 
alloys.\cite{SodaMatsuuraNagaoka,MullerHartmanZittartz,Matsuura}
In normal metals, the ground state of 
the magnetic impurity is known to be a spin singlet,
as a result of the antiferromagnetic coupling 
with the conduction electrons.\cite{Hewson}
However, in the presence of the SC long-range order, 
the energy gap of the host material disturbs 
the conduction electron screening of the local  moment of 
the magnetic impurities.
It was shown in early years that 
whether the local moment is screened  or not 
is determined by the ratio of the Kondo temperature $T_K$ 
to the SC gap $\Delta$. 
The ground state becomes a doublet for $ \Delta \gg T_K $, 
but it is still a singlet for $ \Delta \ll T_K$.
In contrast to the normal metals, for which 
the magnetic solution of the mean-field theory\cite{PWA}
does not describe the correct ground state, 
in the superconducting case the bound state  appearing  
in the energy gap causes 
the occurrence of a quantum phase transition 
to a magnetic doublet ground state.

These aspects of the Kondo physics in superconductors 
have been re-examined precisely by efficient numerical methods 
such as the quantum Monte Carlo (QMC) \cite{JarrelSiviaPatton}
and Wilson numerical renormalization group (NRG) approaches. 
\cite{SSSS,SSSS_2,YoshiokaOhashi,MatsumotoKoga} 
Furthermore, it has also been reconsidered 
for novel systems such as quantum dots.\cite{GlazmanMatveev,Beenakker}
One of the new features in the quantum dot coupled to two SC leads 
is that the ground-state properties of the system can be controlled by  
the phase difference of the two SC gaps $\phi$.   
It induces the Josephson current flowing through the dot,
and at the same time it affects the screening of the local moment. 
When the local moment remains unscreened, 
the spin-flip tunnelling through the magnetic impurity  
causes a current flowing in the opposite direction 
to that in the case of the normal junctions.
An explanation has been given based on the lowest-order 
perturbation theory with respect to the spin-flip tunnelling Hamiltonian 
or equivalently treating the magnetic moment to be a classical spin.
\cite{Kulik,ShibaSoda,SpivakKivelson} 
However, in the Kondo regime, 
the quantum-mechanical nature of the moment is essential 
to the screening.
Therefore, in order to clarify 
how the Josephson current and the dynamics of the moment affect each other, 
the higher order terms in the tunnelling matrix element  
must be taken into account. 
So far, several studies have been carried out using   
the noncrossing approximation (NCA),\cite{Ishizaka,ClerkAmbegaokar} 
slave-boson mean-field (SBMF) theory,\cite{RozhkovArovas,Avishiai}
perturbation theory in $U$,\cite{Vecino}
QMC,\cite{Kusakabe} and NRG.\cite{Choi}
Rozhkov and Arovas
compared the ground-state energy for $\phi=0$ 
and $\phi=\pi$ to discuss the phase transition 
from  the 0 to $\pi$ phase.\cite{RozhkovArovas} 
Clerk and Ambegaoka
pointed out that this transition can be explained 
in terms of the bound state appearing in the SC gap,\cite{ClerkAmbegaokar}
as in the case of the magnetic impurity in the bulk superconductors.
\cite{SodaMatsuuraNagaoka,MullerHartmanZittartz,Matsuura}
So far, mainly one special case,
 in which the absolute value of the two 
SC gaps are the same $|\Delta_L| = |\Delta_R|$ 
and the couplings to the two leads are 
equal $\Gamma_L = \Gamma_R$, 
has been examined.\cite{Ishizaka,ClerkAmbegaokar,RozhkovArovas,Avishiai,Vecino,Choi}
However, the asymmetry in the two gaps $|\Delta_L| \neq |\Delta_R|$ 
and that in the mixing parameters
 $\Gamma_L \neq \Gamma_R$ 
affect sensitively the SC proximity effects 
on the impurity site.  
The superconducting correlations  penetrating from the two leads  
 determine the effective field that is coupled directly to the impurity. 
 Therefore, the asymmetries in the gaps and in the mixing matrix elements,
 as well as the phase difference $\phi$, 
  will also affect the bound state 
 and low-temperature properties of the impurity.

The purpose of this work is to study  
the interplay of the Josephson current 
and the dynamics of the local moment.
To this end,
we start with the Anderson impurity that is connected to two 
SC leads. Although it is already a simplified model,
it still contains a number of parameters.
In the present paper we mainly consider 
the asymmetric gap of $|\Delta_L| \gg |\Delta_R|$, 
where one of the SC gaps is much larger than the others. 
We show that in the limit of $|\Delta_L| \to \infty$ 
the starting model can be mapped onto 
a single-channel model, 
in which the left lead is chopped off 
leaving an extra onsite SC gap of 
$\Delta_d \equiv \Gamma_L \, \text{e}^{\text{i} \theta_L}$ 
at the impurity site. 
This single-channel model possesses both the Josephson and Kondo aspects
since the phase difference can be defined with respect to 
$\Delta_d$ and $\Delta_R$, 
and the simplification gives us a great advantage 
in carrying out numerical calculations. 
Based on this single-channel model,
we examine the ground-state properties for $|\Delta_L| \gg |\Delta_R|$ 
using the NRG method. The phase diagram of 
the ground state is calculated for a wide parameter range of 
the Coulomb interaction $U$ and the mixing strength $\Gamma_R$.
The results show that the phase difference $\phi$ 
tends to make the stability of the singlet ground state worse,
and it can hold a casting vote to determine 
whether the ground state is a singlet or doublet 
when the other parameters are competing.  
Specifically, at $\phi \simeq \pi$, $\Gamma_L 
\simeq \Gamma_R$, 
and near half-filling, the bound state lies close to the Fermi level, 
and it causes a re-entrant behavior of the magnetic doublet ground state.

In \S \ref{sec:Model}, the starting model and 
the mapping onto the single-channel model are described.
In \S \ref{sec:bound_state}, 
 the bound state in the noninteracting case is discussed briefly.
In \S \ref{sec:results} 
the results of NRG calculations are presented. 
Summary is given in \S \ref{sec:summary}.
In Appendix, the  perturbation theory 
for a zero-energy bound state 
and other related matters are provided.

\section{Mapping onto a single-channel Model}
\label{sec:Model}

In this section we introduce the single-channel model, which 
captures the essence of both the Josephson and Kondo physics, 
starting from the model consisting of a single impurity and two SC leads.

The Hamiltonian of the Anderson impurity connected to 
two superconducting leads, at left ($L$) and right ($R$), 
is given by
\begin{align}
\mathcal{H}  
& \, \equiv \,  
      {\cal H}_{d}^0 
      \, + \, {\cal H}_{d}^{U} 
\,+ \, \sum_{\lambda=L,R} \left(\, {\cal H}_{\lambda} 
 \,+ \, {\cal H}_{d\lambda}^{T} \,\right)
\label{eq:H}
\;,
\end{align}
where ${\cal H}_{d}^0$ and ${\cal H}_{d}^{U}$ 
represent the impurity part,
and $\lambda = L,\,R$ is the label assigned for the SC leads. 
The explicit form of each  part is given by 
\begin{align}
{\cal H}_{d}^0 
&\, =
\left(\epsilon_d + \frac{U}{2} \right)  \left( n_{d} -1 \right) \;, 
\quad
{\cal H}_{d}^{U} 
= 
\frac{U}{2} \left(  n_{d} -1 \right)^2 ,
\label{eq:H_int}
\\
{\cal H}_{\lambda} 
&\, =\,
\sum_{k\sigma} 
\epsilon_{\lambda k}^{\phantom{0}}
\,c^{\dagger}_{\lambda k \sigma} 
c^{\phantom{\dagger}}_{\lambda k \sigma} 
\nonumber \\
& \ \ +    
\sum_{k}  
\left(\,
\Delta_{\lambda}\,
c^{\dagger}_{\lambda\,k\,\uparrow}\,
c^{\dagger}_{\lambda\,-k\,\downarrow} \, + \,  
\Delta_{\lambda}^*\,
c^{\phantom{\dagger}}_{\lambda\,-k\,\downarrow}
c^{\phantom{\dagger}}_{\lambda\,k\,\uparrow}\,
\right) ,
\label{eq:sum_within_Debye}
\\
{\cal H}_{d\lambda}^{T} 
&\,=\, -\, 
\sum_{\sigma} 
v_{\lambda}^{\phantom 0}  
\left(\,
 c^{\dagger}_{\lambda \sigma} \, d^{\phantom{\dagger}}_{\sigma} 
\, + \,  
 d^{\dagger}_{\sigma} \,
 c^{\phantom{\dagger}}_{\lambda \sigma}  
\, \right)
\;.
\end{align}
The operator $d^{\dagger}_{\sigma}$ 
 creates an electron with spin $\sigma$ at 
the dot, and  $n_{d} = \sum_{\sigma}
d^{\dagger}_{\sigma} d^{\phantom{\dagger}}_{\sigma}$.  
Similarly, $c^{\dagger}_{\lambda k \sigma}$ 
creates a conduction electron in the lead at $\lambda$,
and $\epsilon_{\lambda k}$ and  $\Delta_{\lambda}$ are 
the energy and SC gap for the electrons in the lead, respectively.
The tunnelling Hamiltonian ${\cal H}_{d\lambda}^{T}$
 connects the dot at the center and lead at $\lambda$. 
At the interface the local operator of the conduction electrons
is defined by 
 $c^{\phantom{\dagger}}_{\lambda \sigma} = \sum_k
c^{\phantom{\dagger}}_{\lambda k \sigma} / \sqrt{N}$. 
Correspondingly, 
 the current flowing into the dot from the left lead $J_L$, 
and the current flowing out to the right lead  $J_R$,
are given by 
\begin{align}
 J_L  &\,=\, \text{i}\,     
 e v_{L}^{\phantom{\dagger}} 
\sum_{\sigma}
           \left(\, 
            d^{\dagger}_{\sigma} 
            c^{\phantom{\dagger}}_{L \sigma}  
          - c^{\dagger}_{L \sigma} 
            d^{\phantom{\dagger}}_{\sigma}
\, \right) 
\label{eq:J_L}
,\\
 J_R  &\,=\, \text{i}\,     
 e v_{R}^{\phantom{\dagger}}
\sum_{\sigma}
  \left(\, 
           c^{\dagger}_{R \sigma} 
           d^{\phantom{\dagger}}_{\sigma} 
        -  d^{\dagger}_{\sigma} 
           c^{\phantom{\dagger}}_{R  \sigma} 
\, \right) \;.
\label{eq:J_R}
\end{align}
Note that we are using units $\hbar=1$ unless otherwise noted.

To discuss the superconducting properties,
we use $2\times 2$ matrix Green's function of the Nambu formulation,
\begin{equation}
\mbox{\boldmath $G$}_{dd}(\tau)
\,=\, -\, 
 \left[ \,
 \begin{matrix}
\langle T_{\tau} \, 
  d_{\uparrow}^{\phantom{\dagger}}(\tau)\, d_{\uparrow}^{\dagger}
\rangle 
&
\langle T_{\tau} \, 
  d_{\uparrow}^{\phantom{\dagger}}(\tau)\, d_{\downarrow}^{\phantom{\dagger}}
\rangle 
\cr
\langle T_{\tau} \, 
  d_{\downarrow}^{\dagger}(\tau) \, d_{\uparrow}^{\dagger}
\rangle 
& 
\langle T_{\tau} \, 
 d_{\downarrow}^{\dagger}(\tau) \, d_{\downarrow}^{\phantom{\dagger}}
\rangle 
 \end{matrix}
\, \right]  \;.
\label{eq:Green_def}
\end{equation}
The Fourier transform of the impurity Green's function, 
$
\mbox{\boldmath $G$}_{dd}(\text{i} \omega_n) 
= 
\int_0^{\beta} \!\! \text{d}\tau\, \text{e}^{\text{i}\omega_n \tau} \,
\mbox{\boldmath $G$}_{dd}(\tau)
$, can be expressed as
\begin{align}
& 
\!\!\!\!\!
\left\{ \mbox{\boldmath $G$}_{dd}(\text{i} \omega_n) \right\}^{-1} 
 \,=\, \text{i} \omega_n \mbox{\boldmath $1$} 
\,- \,E_d\, \mbox{\boldmath $\tau$}_3 
\,-\, v_L^{2}\,
\mbox{\boldmath $\tau$}_3 \,
\mbox{\boldmath $g$}_{L}^{\phantom{0}}(\text{i} \omega_n)\,
\mbox{\boldmath $\tau$}_3
\nonumber
\\
& \qquad \qquad \qquad \quad 
\,- \,v_R^{2} \,
\mbox{\boldmath $\tau$}_3 \,
\mbox{\boldmath $g$}_{R}^{\phantom{0}}(\text{i} \omega_n)\,
\mbox{\boldmath $\tau$}_3 \,-\,
\mbox{\boldmath $\Sigma$}(\text{i} \omega_n)
\;,
\label{eq:Gdd_inv}
\end{align}
where $E_d \equiv \epsilon_d + U/2$,
and 
$\mbox{\boldmath $\tau$}_i$ for $i=1,2,3$ is 
the Pauli Matrix, and  
 $\mbox{\boldmath $\Sigma$}(\text{i} \omega_n)$ is the self-energy 
due to ${\cal H}_d^{U}$.  
Also, the Green's function for the bulk superconductor is given by 
$\left\{ \mbox{\boldmath $g$}_{\lambda k}(\text{i} \omega_n) \right\}^{-1} 
  = 
 \text{i} \omega_n  \mbox{\boldmath $1$} - 
\epsilon_{\lambda k}\,\mbox{\boldmath $\tau$}_3
 - 
\mbox{\boldmath $\Delta$}_{\lambda}$. 
The interface Green's function which enters in Eq.\ (\ref{eq:Gdd_inv})
is defined by
$\mbox{\boldmath $g$}_{\lambda}^{\phantom{0}}(\text{i}\omega_n)
 \equiv \sum_k \mbox{\boldmath $g$}_{\lambda k}(\text{i} \omega_n)/N
$, and can be written as 
\begin{equation}
 \mbox{\boldmath $g$}_{\lambda}^{\phantom{0}}(\text{i}\omega_n)
\, = \,  - \, \pi \rho_{\lambda}^{\phantom{0}}(0) 
\, 
{ \text{i} \omega_n  \mbox{\boldmath $1$} + 
\mbox{\boldmath $\Delta$}_{\lambda}
\over 
\sqrt{\omega_n^2 + |\Delta_{\lambda}|^2 }
} \;,
\label{eq:Glead_2}
\end{equation}
where
the density of states  $\rho_{\lambda}^{\phantom{0}} (\varepsilon) 
= \sum_k \delta(\varepsilon -\epsilon_{\lambda k})/N$ has been 
assumed to be a constant.  
The SC order parameter enters in the off-diagonal element
of $\mbox{\boldmath $\Delta$}_{\lambda}$,
\begin{equation}
\mbox{\boldmath $\Delta$}_{\lambda}
\equiv
\left[ \,
 \begin{matrix}
 0 & \Delta_{\lambda} \cr
 \Delta_{\lambda}^* & 0
 \end{matrix}
 \, \right]  = 
  |\Delta_{\lambda} | \left(\,
  \cos \theta_{\lambda} \,\mbox{\boldmath $\tau$}_1
   - \sin \theta_{\lambda} \,\mbox{\boldmath $\tau$}_2
  \,\right) ,
\end{equation}
where  
  $\Delta_{\lambda} = |\Delta_{\lambda} |\, 
  \text{e}^{\text{i} \theta_{\lambda} }$.
 When the phase difference $\phi \equiv \theta_R -\theta_L$  
 is finite, the Josephson current flows through the impurity.
Furthermore, due to the hybridization with the electrons in the SC leads, 
the superconducting correlation 
$\chi_d \equiv \langle d_{\downarrow} d_{\uparrow} \rangle$, 
which corresponds to the off-diagonal element of 
$\mbox{\boldmath $G$}_{dd}$,
evolves at the impurity site. 

In the noninteracting case, ${\cal H}_{d}^U = 0$,  
the impurity Green's function 
is given by  
\begin{align}
&\left\{ \mbox{\boldmath $G$}_{dd}^0(\text{i} \omega_n) \right\}^{-1} 
= 
\nonumber 
\\ 
& \ \ \ \  \text{i} \omega_n \mbox{\boldmath $1$} 
-  E_d\,\mbox{\boldmath $\tau$}_3 
 +  \Gamma_L  
{ \text{i} \omega_n  \mbox{\boldmath $1$} - 
\mbox{\boldmath $\Delta$}_{L}
\over 
\sqrt{\omega_n^2 + |\Delta_{L}|^2 }
} 
+  \Gamma_R  
{ \text{i} \omega_n  \mbox{\boldmath $1$} - 
\mbox{\boldmath $\Delta$}_{R}
\over 
\sqrt{\omega_n^2 + |\Delta_{R}|^2 }
} ,
\label{eq:Gdd0_inv_explicit}
\end{align}
where $\Gamma_{\lambda}^{\phantom{0}} = 
\pi \rho_{\lambda}^{\phantom{\dagger}}(0) \,v_{\lambda}^2$. 
Correspondingly, the onsite SC correlation and Josephson current 
can be written as
\begin{align}
\chi_{d}^0
& \,=\,    
 \frac{1}{\beta} 
 \sum_{\omega_n}
{ 1\over \det 
\left\{ \mbox{\boldmath $G$}_{dd}^{0}(\text{i}\omega_n) \right\}^{-1} 
 }
 \nonumber \\
& \ \ \ \ \times 
\left[\,
{ \Gamma_L \, 
\Delta_{L}
\over 
\sqrt{\omega_n^2 + |\Delta_{L}|^2 }
} 
\ + \  
{ \Gamma_R \, 
\Delta_{R}
\over 
\sqrt{\omega_n^2 + |\Delta_{R}|^2 }
} 
\,\right] \;,
\label{eq:chi0}
\\
\langle J_R \rangle^0
  & \,=\,   \frac{e}{\beta} 
 \sum_{\omega_n}
{
- 1
\over \det 
\left\{ \mbox{\boldmath $G$}_{dd}^{0}(\text{i}\omega_n) \right\}^{-1} 
 }\, 
\nonumber 
\\
& \ \ \ \ \times 
{ 4\,\Gamma_R\Gamma_L
\,|\Delta_{R}|\,|\Delta_{L}|
\ \sin(\theta_R - \theta_L)
\over 
\sqrt{\omega_n^2 + |\Delta_{R}|^2 }
\sqrt{\omega_n^2 + |\Delta_{L}|^2 }
} 
\;,
\label{eq:J_R_green} 
\end{align}
where $\beta=1/T$, 
and $\langle J_L \rangle^0 =  \langle J_R \rangle^0$.  
The Coulomb interaction $U$ disturbs 
the SC correlation to penetrate into the impurity.
This tendency can be seen already 
in an approximation of the Hartree-Fock level, 
through an anomalous contribution  
\cite{Matsuura,YoshiokaOhashi}
\begin{equation}
{\cal H}_{d}^{U} 
\,\simeq\,
U \left[\ 
   d^{\dagger}_{\uparrow}   d^{\dagger}_{\downarrow}\,
 \langle 
  d^{\phantom{\dagger}}_{\downarrow}  d^{\phantom{\dagger}}_{\uparrow}
 \rangle 
\, + \,
 \langle
  d^{\dagger}_{\uparrow}  d^{\dagger}_{\downarrow}
 \rangle\,
 d^{\phantom{\dagger}}_{\downarrow}   d^{\phantom{\dagger}}_{\uparrow}
\ \right] \ + \  \cdots  \ \;.
\end{equation}
This term corresponds to a Fock term in the Nambu formulation  
\begin{equation}
\mbox{\boldmath $\Sigma$}_{\text{F}}
\,= \,
U\, 
\left[ \,
 \begin{matrix}
 0 &  
\chi_d
\cr
\chi_d^{*}
 & 0
 \end{matrix} 
\,\right]  
\;,
\label{eq:Fock_term}
\end{equation}
and has a different sign from that of the attractive interaction 
which drives the leads  superconducting states. 
However,  to study precisely the effects of the Coulomb interaction 
on the Josephson current and on the SC correlations at the impurity, 
the higher-order terms  beyond the mean-field theory should 
be taken into account. 
To carry out nonperturbative calculations,
the present version of Anderson impurity already contains 
a number of parameters compared to that in the normal leads, 
i.e., we have $|\Delta_{\lambda}^{\phantom{0}}|$ and 
$\theta_{\lambda}$ in addition to $\Gamma_{\lambda}$,  $\epsilon_d$ and $U$.
Therefore, it seems to be meaningful to examine some special cases, 
at which the model can be simplified 
without losing the essence of Josephson and Kondo physics. 
In the following, we consider some such special cases.

\subsection{Model I: $|\Delta_L| = |\Delta_R|$}
\label{subsec:case_I}

 One of the cases that has been studied by 
 a number of authors so far is $|\Delta_L| = |\Delta_R|$  ($\equiv \Delta$), 
 where the absolute value of the two gaps are equal.
 Physically, it means that the two leads are made of 
 the same SC material. 
 In this case the noninteracting Green's function can be written in the form
\begin{align}
& \!\!\!\!\!\!\!\!\!\!\! 
\left\{ \mbox{\boldmath $G$}_{dd}^0(\text{i} \omega_n) \right\}^{-1} 
= \text{i} \omega_n \mbox{\boldmath $1$} 
-  E_d\,\mbox{\boldmath $\tau$}_3 
 +  (\Gamma_L + \Gamma_R )
{ \text{i} \omega_n  \mbox{\boldmath $1$} - 
\widetilde{\mbox{\boldmath $\Delta$}}
\over 
\sqrt{\omega_n^2 + \Delta^2 }
} ,
\label{eq:Gdd0_inv_inversionS}
  \\
\widetilde{\mbox{\boldmath $\Delta$}}
&\,\equiv\, 
\, 
\frac{\Delta}{\Gamma_L + \Gamma_R}
\, \Bigl[\,
\,\left(
\Gamma_L\, \cos \theta_L 
\,+\,
\Gamma_R\, \cos \theta_R 
\right)
\, \mbox{\boldmath $\tau$}_1
\nonumber
\\
& \qquad \qquad \qquad 
\,-\,
\left(
\Gamma_L\, \sin \theta_L 
\,+\,
\Gamma_R\, \sin \theta_R 
\right)
\, \mbox{\boldmath $\tau$}_2
\,\Bigr] ,
\label{eq:Delta_tilde}
\end{align}
and thus we have only one characteristic energy scale 
for the superconductivity, i.e., $\Delta$. 
 However, in this case the two-channel nature of 
 the system is still preserved for finite $\phi$.
This feature can be seen in the functional form 
of $\mbox{\boldmath $G$}_{dd}^0(\text{i} \omega_n)$  
in Eq.\ (\ref{eq:Gdd0_inv_inversionS}).
It becomes equivalent to that of a single-channel model 
only for $\text{e}^{i\theta_R}=\text{e}^{i\theta_L}$, 
when the amplitude of $\widetilde{\mbox{\boldmath $\Delta$}}$ 
coincides with $\Delta$, i.e.,
$
 \widetilde{\mbox{\boldmath $\Delta$}}^2 
 = 
 \Delta^2 \,  \mbox{\boldmath $1$}$.
Therefore, 
in the case of $|\Delta_L| = |\Delta_R|$, 
both of the superconductors must be taken into account 
explicitly to investigate the Josephson effect.

\subsection{Model II: $|\Delta_L| \gg |\Delta_R| $}
\label{subsec:case_II}

The situation is different when one of the gaps is much larger 
than the other's, $|\Delta_{L}| \gg  |\Delta_{R}|$. 
Specifically, in the limit of $|\Delta_{L}| \to \infty$,  
the noninteracting Green's 
function $\mbox{\boldmath $G$}_{dd}^0(\text{i}\omega_n)$ can be written as
\begin{align}
& \!\!\!\! 
\left\{ \mbox{\boldmath $G$}_{dd}^0(\text{i} \omega_n) \right\}^{-1} 
=
\nonumber \\
& \qquad \   \text{i} \omega_n \mbox{\boldmath $1$} 
- \, E_d \, \mbox{\boldmath $\tau$}_3 \,
- \, \Gamma_L \mbox{\boldmath $\tau$}_1
- v_R^{2}\,
\mbox{\boldmath $\tau$}_3 \,
\mbox{\boldmath $g$}_{R}^{\phantom{0}}(\text{i} \omega_n)\,
\mbox{\boldmath $\tau$}_3 
\, .
\label{eq:Gdd_simple}
\end{align}
 where the phase of $\Delta_L$ is chosen to be $\theta_{L} =0$. 
One notable feature is that 
the third term in the right-hand side of Eq.\ (\ref{eq:Gdd_simple}) 
can be regarded as a static superconducting gap induced at the impurity site, 
and its value is given by $\Gamma_L$. 
This term comes from the mixing self-energy due the left lead 
$ v_L^{2}
\mbox{\boldmath $\tau$}_3 
\mbox{\boldmath $g$}_{L}^{\phantom{0}}(\text{i} \omega_n)
\mbox{\boldmath $\tau$}_3
$, and in the limit of $|\Delta_L| \to \infty$  
the off-diagonal SC correlation 
remains finite to penetrate into the impurity site 
while the diagonal damping part corresponding 
to the level width vanishes.
Therefore, 
in the case of  $|\Delta_L| \gg |\Delta_R| $ 
the information about the left lead can be included 
through the off-diagonal $\Gamma_L \mbox{\boldmath $\tau$}_1$ term 
of the impurity Green's function, which can also be  
described by an effective single-channel Hamiltonian  
with an extra SC gap at the impurity site $\Delta_d \equiv \Gamma_L$,
\begin{align}
{\cal H}_{\text{eff}}  
&\,\equiv\,  
      {\cal H}_{d}^0 
      \, + \, {\cal H}_{d}^{U} 
\,+ \, 
{\cal H}_d^{\text{SC}} 
\,+ \, 
{\cal H}_{R} 
 \, + \, {\cal H}_{dR}^{T} 
\;, 
\label{eq:H_eff}
\\
{\cal H}_d^{\text{SC}} &\,=\,
 \Delta_d\,  \left[\,
 d_{\uparrow}^{\dagger} d_{\downarrow}^{\dagger}
\,+\, 
 d_{\downarrow}^{\phantom{\dagger}} d_{\uparrow}^{\phantom{\dagger}}
\,\right]    
\;.
\end{align}
The phase difference $\phi$ defined with respect to the starting model 
is described by the phase difference between $\Delta_R$ and $\Delta_d$.
Using the effective Hamiltonian ${\cal H}_{\text{eff}}$,
one can calculate the expectation values such as 
the Josephson current $\langle J_R \rangle$  
and the Green's function 
$\mbox{\boldmath $G$}_{dd}(\text{i}\omega_n)$ for the interacting electrons,
which coincide with that of the starting model 
in the limit of $|\Delta_L|\to \infty$.
The mapping introduced here is exact. 
It can be confirmed, for instance, using the path integral formulation:
the same effective action for the impurity can be derived 
from both ${\cal H}$ and ${\cal H}_{\text{eff}}$ 
after carrying out the integration over the conduction-electron 
degrees of freedom [see Appendix \ref{sec:action}].
The single-channel Hamiltonian ${\cal H}_{\text{eff}}$ 
can be regarded as a minimal model 
that possesses both the Josephson and Kondo aspects of the system.
This simplification gives us a great advantage 
in carrying out numerical calculations. 
Particularly, the NRG approach works excellently 
for single-channel systems,
while the numerical accuracy becomes rather worse for multi-channel systems 
because the number of the low-lying states 
that should be retained in the calculation 
increases with the number of the channels.
Therefore, in the present study, 
we mainly consider the asymmetric SC gap of $|\Delta_L| \gg |\Delta_R|$ 
using the single-channel Hamiltonian Eq.\ (\ref{eq:H_eff}).

\section{Bound state in the gap}
\label{sec:bound_state}

The bound state appearing in the SC gap plays an important role on 
the ground-state properties of dilute magnetic alloys. 
\cite{SodaMatsuuraNagaoka,MullerHartmanZittartz,Matsuura}
In the case of the quantum dots, 
the Josephson phase $\phi = \theta_R -\theta_L$ also changes the energy 
and wavefunction of the bound state,\cite{ClerkAmbegaokar}
and it also affects the screening of the local moment.
To make these features of the $\phi$ dependence 
of the bound state clear, we discuss briefly about the bound state 
in the noninteracting case [see also Appendix \ref{sec:bogoliubov}].

The bound state can exist in the gap region 
$|\epsilon| < \min (|\Delta_L|, |\Delta_R|)$, 
and the eigenvalue 
is determined by the equation  
$\det \left\{ \mbox{\boldmath $G$}_{dd}^0(\epsilon) \right\}^{-1} =0$.
For noninteracting electrons  
the determinant can be written explicitly 
as,  
\begin{align}
& \det \left\{ \mbox{\boldmath $G$}_{dd}^0(\epsilon) \right\}^{-1} 
= 2 \left[\,  F_2(\epsilon^2)\, -\, F_1(\epsilon^2)  \,\right] \;,
\label{eq:det_ret_inv}
\\
& F_1(\epsilon^2) \ \equiv \ 
\frac{1}{2}  \left[\, \Gamma_L^2 \,+\, \Gamma_R^2
\,+\, E_d^2  
\,-\,\epsilon^2
\,\right] \;,
\label{eq:F_1}
\\
& F_2(\epsilon^2)\ \equiv \ 
{\Gamma_L \, \epsilon^2\,
 \over \sqrt{|\Delta_{L}|^2 - \epsilon^2 } 
 } \,+ \,
{\Gamma_R \, \epsilon^2\,
 \over \sqrt{|\Delta_{R}|^2 - \epsilon^2 } 
 } 
 \nonumber
 \\
 & \qquad \qquad 
 \,+\,
{\Gamma_L \Gamma_R  
\left(\, \epsilon^2\,-\,
|\Delta_{L}| |\Delta_{R}| \, \cos \phi \,\right)
 \over 
 \sqrt{|\Delta_{L}|^2 - \epsilon^2} 
 \sqrt{|\Delta_{R}|^2 - \epsilon^2} 
 } 
 \;.
 \label{eq:F_2}
\end{align}
Here $F_1(\epsilon^2)$ is a simple decreasing function of $\epsilon^2$,
while the functional form of $F_2(\epsilon^2)$ depends 
on the parameters  
$|\Delta_{L}^{\phantom{0}}|$, $|\Delta_{R}^{\phantom{0}}|$ and $\phi$.
However, at $\epsilon=0$, it is given simply by 
$F_2(0) = 
-\, \Gamma_L \Gamma_R\, \cos \phi$.
Therefore $F_1(0) \geq F_2(0)$, 
and the equality holds for
 $\phi=\pi$, $\Gamma_L= \Gamma_R$ and $E_d=0$.
 We consider this particular case in the following, 
 and then discuss  briefly the bound state for
 the models mentioned in \S \ref{sec:Model}.

\begin{figure}[bt]
\begin{center}
\leavevmode
\includegraphics[width=1\linewidth, clip, 
trim = 0.5cm 0.5cm 0.5cm 0.5cm]{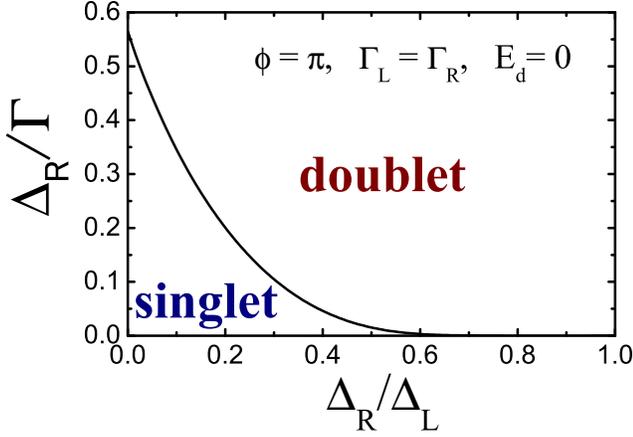}
\caption{
Phase diagram of the ground state 
for $\phi=\pi$, $\Gamma_L=\Gamma_R$ ($\equiv \Gamma$), 
and $E_d=0$. Here  $|\Delta_L|$ is chosen 
to be $|\Delta_L| \geq |\Delta_R|$. 
An infinitesimal $U$
lifts the 4-fold degeneracy caused by the zero mode 
to stabilize the singlet or doublet state depending on 
the value of  $|\Delta_R|/|\Delta_L|$ and $1/\Gamma$.
The results are obtained with the perturbation theory
in Appendix \ref{sec:degenerate_perturbation}. 
}
\label{fig:gam_critical}
\end{center}
\end{figure}

\subsection{Zero mode 
  for $\phi=\pi$, $\Gamma_R=\Gamma_L$ and $E_d=0$
}
\label{subsec:zero_mode}

We consider here the special case where 
the three conditions hold simultaneously; 
$i$)  $\pi$-junction $\phi=\pi$, 
$ii$) equal coupling $\Gamma_L = \Gamma_R$
 ($\equiv \Gamma$), and electron-hole symmetry $E_d=0$. 
In this case $F_2(\epsilon^2)$ is a increasing function of $\epsilon^2$,
and at $\epsilon=0$ it coincides with $F_1(0)$,
so that the bound state comes just on the Fermi energy $\epsilon=0$.

Thus, the ground state has 4-fold degeneracy with respect to   
the occupation of the zero-energy bound state by 
the Bogoliubov quasiparticles 
[see also Appendix \ref{sec:pair_potential_at_E}]. 
The Coulomb interaction ${\cal H}_{d}^{U}$ lifts the degeneracy.
We have calculated the energy shift 
for an infinitesimal positive $U$  using  
the perturbation theory, the outline of which is given  
in Appendix \ref{sec:degenerate_perturbation}.
The ground state is a singlet  
for $\Gamma > \Gamma_{\text{cr}}$, 
while  it is a doublet for $\Gamma < \Gamma_{\text{cr}}$.
The critical value  $\Gamma_{\text{cr}}$ depends 
on the ratio $x \equiv |\Delta_R|/|\Delta_L|$ as shown
in Fig.\ \ref{fig:gam_critical}.
In this figure the phase boundary is plotted in the $x$ vs $1/\Gamma$ plane 
choosing $|\Delta_L|$ to be $|\Delta_L| \geq |\Delta_R|$, 
where the energy is scaled by $|\Delta_R|$.
The critical value $\Gamma_{\text{cr}}$ 
increases with $x$ and diverges exponentially at $x=1$, 
and thus for $|\Delta_L|=|\Delta_R|$ the ground state 
is a doublet independent of the value of $\Gamma$. 
This is because in this particular case 
the SC correlation penetrates from the left lead
and that from the right cancel each other out 
to make the net SC correlation 
at the impurity site $\chi_d^0$ to be zero.   
This example shows that the asymmetry in the two 
gaps $|\Delta_L| \neq |\Delta_R|$ affects 
the screening of the moment crucially when 
the other parameters are set to be highly symmetric.
Note that in the  perturbation theory 
an exchange integral (Fock term) of the form Eq.\ (\ref{eq:chi0}) 
shifts the bound state from the Fermi level and 
favors the singlet state, 
whereas a direct type contribution (Hartree term)
favors the doublet state.

 \subsection{Bound state for Models I \& II }
\label{subsec:bound_state_I_II}

 When the absolute value of the two gaps are the same, 
$|\Delta_{L}| = |\Delta_{R}| \equiv \Delta$,
Eq.\ (\ref{eq:F_2}) simplifies as
\begin{equation}
F_2(\epsilon^2) 
\,=\,
{(\Gamma_L+\Gamma_R) \, \epsilon^2\,
 \over \sqrt{\Delta^2 - \epsilon^2 } 
 } 
\,   - 
\Gamma_L \Gamma_R  
   + 
{
2\,\Gamma_L \Gamma_R  
\,\Delta^2\, \sin^2 \frac{\phi}{2}
 \over   \Delta^2 - \epsilon^2}  .
\end{equation}
Since this is an increasing function of $\epsilon^2$ 
which diverges positively at $\epsilon^2 \to \Delta^2$, 
the bound state exists for any finite  $\Delta$.

For the other model with the asymmetric gap $|\Delta_{L}| \gg |\Delta_{R}|$,  
the function $F_2(\epsilon^2)$ is given by  
\begin{equation}
F_2(\epsilon^2) \, = \, 
\Gamma_R \, 
{ \epsilon^2 \,-\, \Delta_d \, |\Delta_{R}| \, \cos \phi 
 \over 
 \sqrt{|\Delta_{R}|^2 - \epsilon^2} 
 } 
 \;,
\end{equation}
where $\Delta_d \equiv  \Gamma_L$. 
In this case the behavior of 
 $F_2(\epsilon^2)$ at $\epsilon^2 \to  |\Delta_{R}|^2 - 0^+$ 
 depends crucially on the sign of the numerator,
and thus whether or not the bound state exists depends on 
the values of the parameters. 

\section{NRG approach}
\label{sec:results}

In the rest of this paper 
we discuss the ground state properties 
of the Anderson impurity in the limit of $|\Delta_L| \gg |\Delta_R|$, 
where one of the SC gaps is much larger than that of the others, 
based on the single-channel model given by Eq.\ (\ref{eq:H_eff}).
As mentioned in \S \ref{subsec:case_II}, 
this model contains essential aspects of both 
the Josephson and Kondo effects. Also, 
the reduction in the number of the channels 
gives us a practical advantage 
in  the NRG calculations.

\subsection{Method}
\label{subsec:NRGmethod}

Through a standard procedure of 
the logarithmic discretization,\cite{KWW}
the conduction band can be modelled 
by a linear chain with the complex pair 
potential $\Delta_R=|\Delta_R|\,\text{e}^{\text{i} \phi}$,
and a sequence of the NRG Hamiltonian $H_N$ is given by
\begin{align}
H_N =&     \Lambda^{(N-1)/2}  
\left[ 
      {\cal H}_{d}^0 
       + {\cal H}_{d}^{U} 
+  {\cal H}_d^{\text{SC}} 
 +  H_{dR}  +   H_{R}
 \right] ,
\label{eq:H_N}
\\
H_{R}\,= &  \  
D\,{1+1/\Lambda \over 2} \,
\nonumber \\
& \times 
\sum_{n=0}^{N-1} 
\sum_{\sigma}
 \xi_n\, \Lambda^{-n/2} 
\left(\,
  f^{\dagger}_{n+1\,\sigma}\,f^{\phantom{\dagger}}_{n \sigma}
  +  
 f^{\dagger}_{n \sigma}\, f^{\phantom{\dagger}}_{n+1\,\sigma}
 \,\right) 
\nonumber \\
& + \   
\sum_{n=0}^{N} \,
\left(\,
\Delta_R^{\phantom{*}}\,
  f^{\dagger}_{n \uparrow} f^{\dagger}_{n \downarrow}
 \, + \, 
\Delta_R^*\,
 f^{\phantom{\dagger}}_{n \downarrow} f^{\phantom{\dagger}}_{n \uparrow}
\,
\right) \;,
\label{eq:NRG_SC_cond}
 \\
H_{dR} \,= & \   D 
       \sum_{\sigma}
        \,\widetilde{v}_R^{\phantom{0}} 
\left(\,
f^{\dagger}_{0 \sigma} d^{\phantom{\dagger}}_{ \sigma}
         \, + \,  
d^{\dagger}_{ \sigma} f^{\phantom{\dagger}}_{0 \sigma} 
          \,  \right)  \;.
\end{align}
Here  $D$ is the half-width of the conduction band,
and $f_{n\sigma}$ is an operator for electrons in the right lead.
The hopping matrix elements   
 $\xi_n$ and $\widetilde{v}_R^{\phantom{0}}$ are defined by
\begin{align}
\xi_n &\,=\,   
{ 1-1/\Lambda^{n+1}  
\over  \sqrt{1-1/\Lambda^{2n+1}}  \sqrt{1-1/\Lambda^{2n+3}} 
} 
\;,\\
  \widetilde{v}_R^{\phantom{0}}
&\,=\, \sqrt{ \frac{2\,\Gamma_R A_{\Lambda}}{\pi D} }
\;,
\qquad
A_{\Lambda}  \,=\,  \frac{1}{2}\, 
 {1+1/\Lambda \over 1-1/\Lambda }
\,\log \Lambda
\;,
\end{align}
where $A_{\Lambda}\to 1$, in the continuum limit $\Lambda \to 1$.
\cite{KWW,SakaiShimizuKasuya}
The low-lying energy states of ${\cal H}_{\text{eff}}$ 
can be deduced from that of $\Lambda^{-(N-1)/2} H_N$ for large $N$. 
In the presence of the complex SC gap 
$H_N$ does not conserve the charge, 
and the total spin $S$ is the only one quantum number that 
can be used for the block diagonalization. 
In the two special cases, $\phi= 0$ and $\pi$,
the Hamiltonian $H_N$ has an extra U(1) symmetry 
corresponding to the $x$-th component of the axial charge
$
I_x = \sum_{n=-1}^N  (-1)^n\,
(
f^{\dagger}_{n\uparrow} f^{\dagger}_{n\downarrow} + 
f^{\phantom{\dagger}}_{n\downarrow} f^{\phantom{\dagger}}_{n\uparrow} )/2
$ with $f^{\dagger}_{-1\sigma} \equiv d^{\dagger}_{\sigma}$,\cite{KWW} 
 and the SC lead can be transformed 
into a normal lead with a staggered potential.\cite{SSSS}

The ground-state average of the Josephson current 
 can be obtained using the discretized version of Eq.\ (\ref{eq:J_R}), 
\begin{align}
& 
\langle J \rangle_N \ = 
\nonumber 
\\
& \qquad 
\, \text{i}\,     
 e \,\widetilde{v}_{R}^{\phantom{\dagger}} D 
\sum_{\sigma}
  \left( 
  \langle  \Phi_N | 
           f^{\dagger}_{0 \sigma}  
           d^{\phantom{\dagger}}_{\sigma} 
 | \Phi_N \rangle 
        -   
  \langle \Phi_N | 
        d^{\dagger}_{\sigma}  
           f^{\phantom{\dagger}}_{0  \sigma} 
 | \Phi_N  \rangle 
 \right) ,
\label{eq:J_R_NRG} 
\end{align}
where $| \Phi_N \rangle$ is the ground state of $H_N$.
The expectation value can be calculated successively 
for large $N$ using the recursive relation among the matrix elements 
$\langle J \rangle_N$ and $\langle J \rangle_{N+1}$.
 We have carried out the calculations  
 retaining the lowest $500$ states 
and taking $\Lambda$ and $|\Delta_R|$ to be 
$\Lambda=2.0$ and $|\Delta_R|/D=1.0 \times 10^{-5}$. 
In the following,  
we use $|\Delta_R|$ as the unit of the energy in most of the figures, 
since that is a typical energy scale for the superconductivity.

\subsection{Results at half-filling}
\label{subsec:sym}

\begin{figure}[bt]
\begin{center}
\leavevmode
\includegraphics[width=0.95\linewidth, clip, 
trim = 0.5cm 0.6cm 0.5cm 0.6cm]{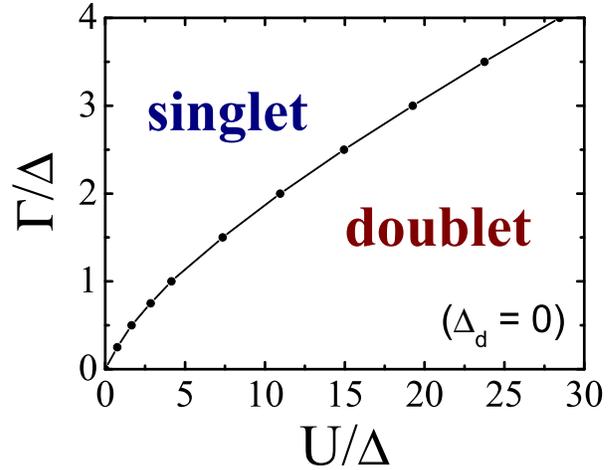}
\caption{
Phase diagram of the ground state 
for the symmetric Anderson model $\epsilon_d = -U/2$
connected to the single SC lead of $\Delta$,
where the SC gap in the impurity site is taken to be $\Delta_d=0$.
The NRG calculations have been carried out 
for $\Lambda = 2.0$  and $\Delta/D=1.0 \times 10^{-5}$. 
}
\label{fig:phase_diagram_1ch}
\end{center}
\end{figure}

In this subsection we show the results obtained 
in the electron-hole symmetric case, $\epsilon_d = - U/2$. 
We first of all consider the ground state for $\Delta_d=0$.
Since the value of $\Delta_d$ is defined to be equal 
to $\Gamma_L$ in the original two-channel model Eq.\ (\ref{eq:H}),
the condition $\Delta_d=0$ means that 
the left superconductor is disconnected completely from the impurity. 
In Fig.\ \ref{fig:phase_diagram_1ch}, 
the phase diagram  of the ground state is  
shown as a function of $U/|\Delta_R|$ and $\Gamma_R/|\Delta_R|$.
The ground state is a spin-singlet state for large $\Gamma_R$ or small $U$, 
and it is a doublet for small $\Gamma_R$ or large $U$.
We note that this particular case, $\Delta_d=0$, has been examined  
by Yoshioka and Ohashi,\cite{YoshiokaOhashi}
and the phase boundary is determined basically 
by a single parameter $T_K/|\Delta_R|$ in the Kondo regime. 
Nevertheless, for small $U$ or the electron-hole asymmetric case, 
in which the contributions of the charge excitations still remain, 
the low-energy properties are characterized  
not only by the single parameter $T_K$, 
but the Wilson ratio $R$ and 
the renormalized impurity level $\widetilde{\epsilon}_d$ 
are necessary for a complete description.\cite{Hewson,Hewson_renorm}
Therefore, in order to see an overall picture of 
the ground-state properties,
the phase diagrams which are plotted as a function of  
the bare parameters, as Fig.\ \ref{fig:phase_diagram_1ch}, 
are convenient.

\begin{figure}[tb]
 \begin{center}
 \leavevmode
\includegraphics[width=0.9\linewidth]{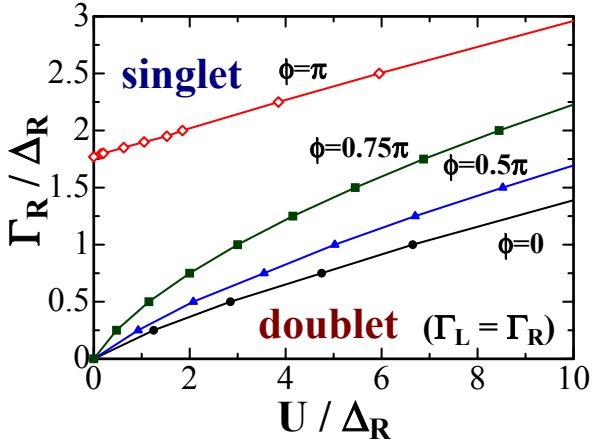}
 \caption{
 Phase diagram of the ground state for 
 the equal coupling $\Gamma_L=\Gamma_R$ at 
 half-filling  $\epsilon_d = -U/2$ 
 is plotted  
 for several values of the Josephson phase $\phi$.
 Here $\Delta_d$ ($\equiv \Gamma_L$) 
 is taken to be equal to $\Gamma_R$.
 For $\phi=\pi$ and  $U=0$, 
 the phase boundary is given by 
 $\Gamma_{\text{cr}} =1.77064 |\Delta_R|$. 
 This special behavior at $\phi=\pi$ is caused by the zero mode, 
 and  an infinitesimal $U$ stabilizes 
 the singlet (doublet) ground state 
 for $\Gamma_R > \Gamma_{\text{cr}}\,$  
 ($\Gamma_R < \Gamma_{\text{cr}}$).   
 }
 \label{fig:equal_couple}
 \end{center}
 \end{figure}

When the SC gap of the impurity  $\Delta_d$ ($\equiv \Gamma_L$) 
becomes finite owing to the coupling with the left lead 
of $|\Delta_L| \to \infty$, the phase difference $\phi$ 
can be introduced at the interface of $\Delta_d$ and $\Delta_R$.
We now examine how $\phi$ affects 
the ground state in the equal-coupling case $\Gamma_L = \Gamma_R$. 
In Fig.\ \ref{fig:equal_couple} 
the phase diagram of the ground state 
is plotted as a function of $U$ and $\Gamma_R$  
for several value of the phase difference 
 $\phi=0,\, \pi/2,\, 3\pi/4,\, \pi$.
Here $\Delta_d$ is varied simultaneously with $\Gamma_R$ 
keeping the relation $\Delta_d = \Gamma_R$.
We see that  $\phi$ tends 
to disturb the screening of the local moment and enlarges 
the doublet phase in the figure.
The feature of the phase diagram at $\phi \neq \pi$  
is similar to that for $\Delta_d=0$ in 
Fig.\ \ref{fig:phase_diagram_1ch}. 
However, 
the feature is quite different for $\phi = \pi$. 
This is caused by the bound state lying just 
on the Fermi level at $U=0$. 
An infinitesimal $U$ lifts the degeneracy to stabilize 
the singlet state at $\Gamma_R > \Gamma_{\text{cr}}$, 
and doublet state at $\Gamma_R < \Gamma_{\text{cr}}$. 
In the present model, $\Delta_L \to \infty$, 
the critical value is given by 
$\Gamma_{\text{cr}} = 1.77064 |\Delta_R|$ as  
described in Appendix  \ref{sec:degenerate_perturbation}.
For this reason, in Fig.\  \ref{fig:equal_couple} 
the phase boundary for $\phi=\pi$ starts from the 
finite value, $\Gamma_{\text{cr}}$, in the $y$ axis.

\begin{figure}[tb]
\begin{center}
\leavevmode
\includegraphics[ width=1\linewidth, clip, 
trim = 0.3cm 0.6cm 0.4cm 0.4cm]{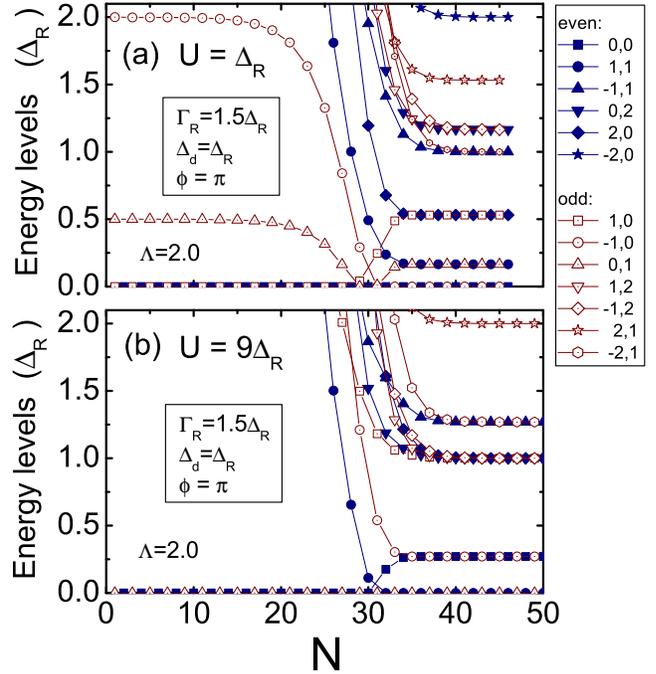}
\caption{Low-lying energy levels of
$\Lambda^{-(N-1)/2} H_N$  as a function of $N$, 
where the energy is scaled by $|\Delta_R|$ and 
is measured from the ground-state energy at each $N$.
The solid and open symbols are the levels for even and odd $N$,
respectively. 
Only the lowest eigenvalue in  each subspace labeled by 
the quantum numbers ($I_x$, $2S$) are shown, where 
$I_x$ is the U(1)-axial charge and $S$ is the total spin.
The parameters are taken to be 
 $\phi = \pi$, $\Delta_d =|\Delta_R|$,  $\Gamma_R= 1.5|\Delta_R|$, 
 $\epsilon_d = -U/2$, and  $\Lambda=2.0$.    
At large $N$, the levels approach to the fixed point values 
showing no even-odd dependence on $N$. 
For $N \gtrsim 32$, the ground state is a singlet in (a) $U=|\Delta_R|$,
 while it is a doublet in (b) $U=9|\Delta_R|$.
}
\label{fig:energy_flow1}
\end{center}
\end{figure}

To see how the singlet or doublet ground states evolves 
in the successive NRG steps, in Fig.\ \ref{fig:energy_flow1} 
the low-lying  energy levels of $\Lambda^{-(N-1)/2} H_N$  
  is plotted  as a function of $N$ 
for  (a) $U=|\Delta_R|$, and (b) $U=9.0|\Delta_R|$.
Here the factor $\Lambda^{-(N-1)/2}$ restores the original energy scale 
of ${\cal H}_{\text{eff}}$, 
and the ground-state energy is subtracted from the eigenvalue
for the excited states.  
 The  parameters are taken to be 
 $\phi=\pi$ and  $\Gamma_R = 1.5|\Delta_R|$. 
 Specifically, the SC gap in the impurity site 
 is chosen to be $\Delta_d = |\Delta_R|$, 
 so that the couplings are asymmetric $\Gamma_L \neq \Gamma_R$ in 
 this parameter set. 
In the figure the eigenstates are labeled by  
the two quantum numbers ($I_x$, $2S$), 
where $I_x$ is the U(1)-axial charge mentioned in the above 
and $S$ is the total spin.
The solid and open symbols correspond to the levels for even and odd $N$,
respectively. 
For large $N$, the excitation energies converge to the fixed point values, 
and have no even-odd oscillatory dependence on $N$,
which is a typical behavior seen in the presence of 
the energy gap.\cite{SSSS}
The ground state is a singlet for $N \gtrsim 32$ in (a), 
while it is a doublet in the case of (b).
In the figures  all the low-lying levels are shown for  $E<|\Delta_R|$,
but not all are shown for $E>|\Delta_R|$. 
The levels below the gap $|\Delta_R|$ can be classified 
according to the occupation of the bound state with 
the Bogoliubov quasiparticles,
which can be inferred from the value of $I_x$ because 
it is transformed into the usual charge 
by the Unitary transformation described in Ref.\ 7. 
In the case of (a), where $U$ is relatively small, 
the first and second excited states lie in the gap region. 
The occupation of the bound state 
in the ground, first, and second excited states 
are supposed to be empty, single, and double, respectively.
In contrast, in (b) the occupation of the bound state in the ground 
and first excited states are supposed to be single and empty, respectively.
The doubly occupied state could not stay below the gap in (b) 
because of large $U$.  

\begin{figure}[tb]
\begin{center}
\leavevmode
\includegraphics[width=0.9\linewidth]{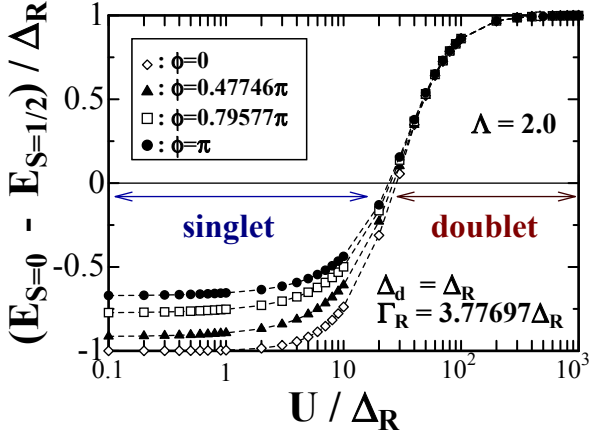}
\caption{
The fixed point value of the lowest excitation 
energy of $\Lambda^{-(N-1)/2} H_N$  at large $N$ 
as a function of $U$.
Here $E_{\text{S=0}}$ and $E_{\text{S=1/2}}$
are the eigenvalue of the lowest singlet and double states, respectively. 
Therefore, the ground state is a singlet (doublet)
for negative (positive) $E_{\rm S=0} - E_{\rm S=1/2}$. 
The parameters are taken to be 
$\Delta_d =|\Delta_R|$,
$\Gamma_R=3.77697|\Delta_R|$, 
 $\epsilon_d = -U/2$, 
and
 $\Lambda=2.0$.
 }
\label{fig:fixed_point}
\end{center}
\end{figure}

The eigenvalues depend on the phase difference $\phi$,
and particularly the fixed-point energy at large $N$   
determines the low-energy properties. 
In Fig.\ \ref{fig:fixed_point},
the fixed-point value of the lowest excitation energy 
 $E_{S=0} - E_{S=1/2}$ is plotted as a function of $U$ for 
 several values of $\phi$, 
where  $E_{S=0}$ and $E_{S=1/2}$ are the lowest eigenvalues 
of $\Lambda^{-(N-1)/2}\,H_N$ in the singlet and 
doublet subspace, respectively.  
The parameters are chosen to be $\Gamma_R = 3.77697|\Delta_R|$, 
and $\Delta_d = \Delta_R$.
For  $U < U_C$ with $U_C \simeq 25 |\Delta_R|$, 
the ground state is a singlet, and
the excitation energy depends visibly on $\phi$.
The energy separation between the two states decreases 
with increasing $\phi$, and it means that
$\phi$ tends to destabilize the singlet ground state.
On the other hand the ground state is a doublet for $U > U_C$, 
and the $\phi$ dependence of the excitation energy is very weak.
This seems to be caused by the fact that 
the local moment, which remains in the doublet ground state, 
makes the SC correlation length small, and 
the phase coherence between the dot and SC lead becomes weak.  
In the singlet state, however, the phase coherence is preserved, 
and thus the excitation energy depends sensitively on $\phi$. 
We can also see in Fig.\ \ref{fig:fixed_point}
that the critical value $U_C$ decreases with increasing $\phi$ to 
reduce the area of the singlet ground state in the phase diagram.

\begin{figure}[tb]
\begin{center}
\leavevmode
\includegraphics[width=0.9\linewidth]{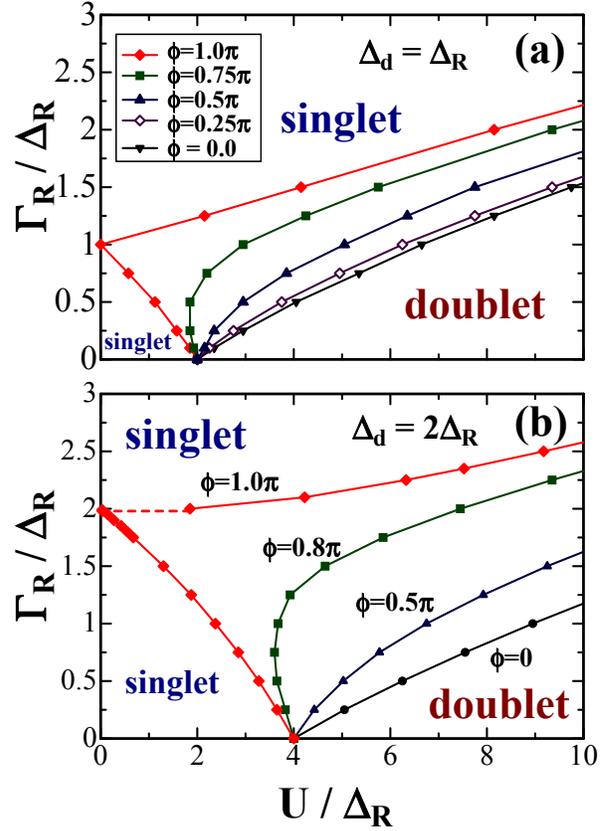}
\caption{
Phase diagrams of the ground state 
in the electron-hole symmetric case  $\epsilon_d = -U/2$.
Here the impurity SC gap is taken to be 
(a) $\Delta_d =|\Delta_R|$,  and  (b) $\Delta_d = 2 |\Delta_R|$.  
At $\Gamma_R=0$, the phase boundary is  
given by $U= 2\Delta_d$ from the atomic-limit solution. 
 For $\phi = \pi$, the bound state comes just on the Fermi level  
 when $U=0$ and $\Gamma_R=\Delta_d$ ($\equiv \Gamma_L$).
In the case of (b), 
this happens in the region 
of $\Gamma_R > \Gamma_{\text{cr}} =1.77064 |\Delta_R|$, 
and on the dotted horizontal line 
the ground state is a singlet.
}
\label{fig:Phase_diagram_josephson}
\end{center}
\end{figure}

 For  asymmetric couplings $\Gamma_R \neq \Gamma_L$, 
the phase diagram of the ground state 
has some different features from that of the equal 
coupling case seen in Fig.\ \ref{fig:equal_couple}. 
Two typical examples for the asymmetric coupling  
are shown in Fig.\ \ref{fig:Phase_diagram_josephson} 
for several values of $\phi$. 
Here the value of $\Delta_d$ is fixed to be 
(a) $|\Delta_R|$ and (b) $2|\Delta_R|$ independent 
of the coupling $\Gamma_R$.
In each of the figures (a) and (b), all the lines 
start from the same point on the horizontal line,
the position of which is given by $U= 2\Delta_d$ 
from the atomic-limit solution at $\Gamma_R=0$. 
 A re-entrant behavior seen for $\phi \simeq \pi$ is 
caused by the bound state lying near the Fermi level.
The phase boundary for $\phi = \pi$  
crosses the vertical axis at $\Gamma_R=\Delta_d$,
at which the couplings becomes equal $\Gamma_L=\Gamma_R$ 
and the bound state comes just on the Fermi level.
An infinitesimal $U$ lifts the 4-fold degeneracy 
of the ground state as shown in Fig.\ \ref{fig:gam_critical},  
and specifically in the present case,  $\Delta_L \to \infty$,  
the critical value is given by  
$\Gamma_{\text{cr}}=1.77064 |\Delta_R|$ 
(see Appendix \ref{sec:degenerate_perturbation}).
Thus  the examples (a) and (b) represent typical  
 cases of $\Delta_d < \Gamma_{\text{cr}}$ 
and $\Delta_d > \Gamma_{\text{cr}}$, respectively. 
In (b),  
the ground state is a singlet 
on the dotted horizontal line starting from $\Gamma_R=\Delta_d$,
and the transition to the doublet state occurs discontinuously 
at finite $U$ at the end of the dotted line.
These examples show that 
the quantum phase transition can be driven by $\phi$ 
if the other parameters are in the region surrounded by 
the lines for $\phi=0$ and  that for $\phi=\pi$. 

\begin{figure}[tb]
\begin{center}
\leavevmode
\includegraphics[width=0.9\linewidth]{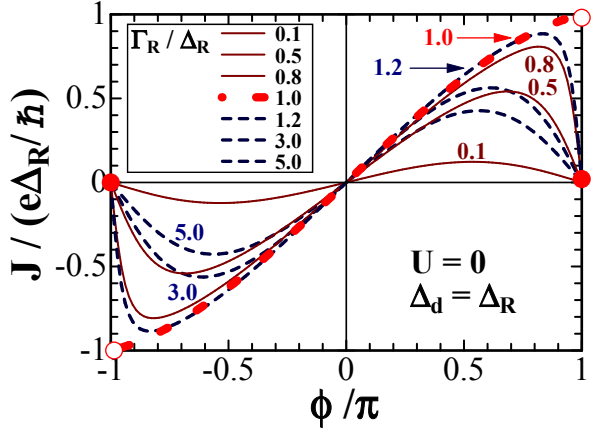}
\caption{
The Josephson current for $U=0$ and $E_d=0$ as a function of $\phi$.
The currents are calculated for several values of $\Gamma_R$ 
from Eq.\ (\ref{eq:J_R_green}).  
Here $\Delta_d$ ($\equiv \Gamma_L$) is taken to be $\Delta_d = |\Delta_R|$, 
so that $\Gamma_R$ becomes equal to $\Gamma_L$ 
for  $\Gamma_R/|\Delta_R| = 1.0$, where 
the current for $\phi = \pm \pi$ jumps discontinuously. 
}
\label{fig:current_for_U0}
\end{center}
\end{figure}

\begin{figure}[bt]
\begin{center}
\leavevmode
\includegraphics[width=0.9\linewidth]{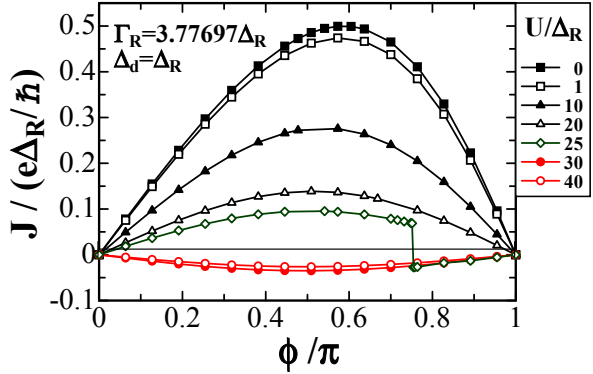}
\caption{
The Josephson current as a function of $\phi$
for various values of $U$. 
Here the  parameters are taken to be  
$\Gamma_R=3.77697\Delta_R$, 
 $\Delta_d =\Delta_R$, 
and $\epsilon_d = -U/2$. 
The ground state is a singlet 
for $U \lesssim 20 \Delta_R$,
while it is a doublet for for $U \gtrsim 30 |\Delta_R|$.
For $U = 25 |\Delta_R|$, the quantum phase transition 
to the doublet ground state occurs at $\phi \simeq 0.76 \pi$,
and the direction of the current changes 
due to the spin-flip tunneling \cite{Kulik}.
Note that the current for negative $\phi$ is symmetric around the origin 
as that in Fig.\ \ref{fig:current_for_U0},
since the current is an odd function of $\phi$ 
with the period $2\pi$.
}
\label{fig:current_for_U}
\end{center}
\end{figure}

We next study how the Coulomb interaction affects the Josephson current.
To this end, we start with noninteracting case.
In Fig.\  \ref{fig:current_for_U0}
the Josephson current for $U=0$, 
which is obtained from Eq.\ (\ref{eq:J_R_green}),
is plotted as a function of $\phi$ for several values of $\Gamma_R$.
Here the parameters are taken to be 
 $\Delta_d = |\Delta_R|$, 
 and thus at $\Gamma_R= |\Delta_R|$ the two couplings 
of the original model become equal $\Gamma_R=\Gamma_L$ 
since $\Delta_d \equiv \Gamma_L$.
When the two couplings are close $\Gamma_R \simeq \Gamma_L$,  
the amplitude of the current becomes large,
and the shape of the $\phi$ dependence deviates 
from a simple sinusoidal form.
In the figure the dotted line is the result for $\Gamma_R=\Gamma_L$,   
and it shows a discontinuous jump at $\phi= \pm \pi$. 
This singularity is caused by the bound state at Fermi level.
The NRG results of the Josephson current are shown 
in Fig.\ \ref{fig:current_for_U} 
for several values of $U$, where
the parameters are taken to be 
$\Gamma_R = 3.77697 |\Delta_R|$, and $\Delta_d = |\Delta_R|$.
In this figure the results 
are plotted only for $0 \leq \phi \leq \pi$. 
The feature of the current at negative $\phi$ 
can be seen by rotating the figure around the origin, 
since the current is an odd function of $\phi$ with the period $2\pi$.
 For small $U$,
 the ground state is a singlet for the whole range 
 of $\phi$ as seen in the results for $U \lesssim 20 |\Delta_R|$,
 and in these cases the amplitude of the current decreases 
 with increasing $U$.
We see for $U = 25 |\Delta_R|$ that 
the phase transition occurs at $\phi \simeq 0.76 \pi$. 
Then in the doublet state the current flows in the opposite 
direction from that in the singlet state. 
This change is caused by 
the spin-flip tunneling through the unscreened 
local moment.\cite{Kulik,ShibaSoda}
Therefore,  the Josephson current can be 
used to detect the quantum phase transition.

\begin{figure}[tb]
\begin{center}
\leavevmode
\includegraphics[width=0.9\linewidth]{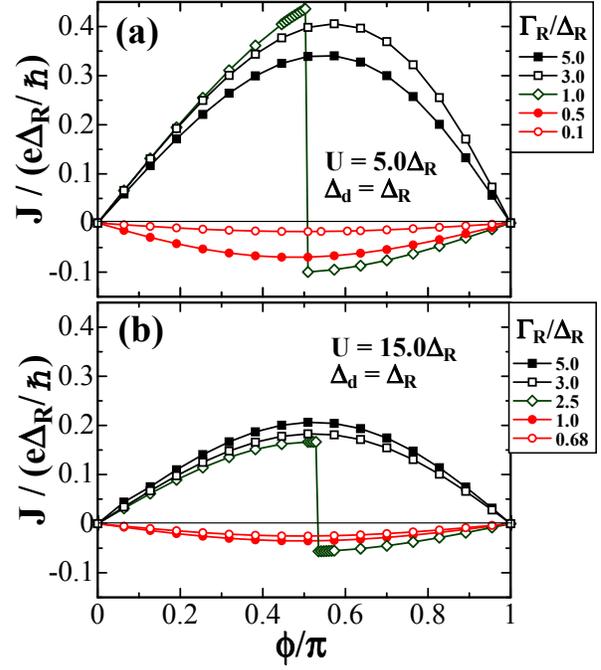}
\caption{
The Josephson current as a function of $\phi$
for various values of $\Gamma_R$: the onsite Coulomb repulsion 
is taken to be 
 (a) $U=5.0|\Delta_R|$ and (b) $U=15.0|\Delta_R|$. 
Other parameters are $\Delta_d =|\Delta_R|$, and 
 $\epsilon_d = -U/2$. 
}
\label{fig:current_for_Gamma}
\end{center}
\end{figure}

In quantum dots the coupling to the leads is  
a tunable parameter. 
We have plotted the current as a function of $\phi$ 
in Fig.\ \ref{fig:current_for_Gamma}  
  for several values of $\Gamma_R$ 
 taking $U$ to be  (a) $5.0|\Delta_R|$ and (b) $15.0|\Delta_R|$. 
Here the gap is chosen to be $\Delta_d = |\Delta_R|$. 
Both in (a) and (b), 
the phase transition from the singlet state to doublet state occurs
when $\Gamma_R$ decreases. 
It can be explained from the fact that the correlation effects 
are enhanced for small $\Gamma_R$.
However, the amplitude of the current depends on another factor.
In Fig.\ \ref{fig:current_for_Gamma} (a), 
the amplitude becomes large when the value 
of $\Gamma_R$ approaches  $\Gamma_L$ ($\equiv \Delta_d$).
In this case, the matching condition at the interface 
is the dominant factor that determines the amplitude of the current.
On the other hand, the amplitude decreases with $\Gamma_R$ 
in Fig.\ \ref{fig:current_for_Gamma} (b). 
In this case, the ground state is a doublet 
when $\Gamma_R$ becomes equal to $\Gamma_L$,
and thus in the singlet state at $\Gamma_R >\Gamma_L$ 
the strong electron correlation dominates the matching condition 
to make the amplitude of the current small.

\subsection{Results away from half-filling}
\label{subsec:asym}

So far, we have discussed the results at half-filling $\epsilon_d= -U/2$, 
where the impurity site is singly occupied in average.
In that case, the magnetic correlations are enhanced,
while the charge excitations are suppressed. 
Therefore, when the system goes away from half-filling,
the magnetic doublet ground state should be destabilized. 
In quantum dots the impurity potential $\epsilon_d$ is the
parameter that can be controlled by the gate voltages.
We show in the following the NRG results for the ground-state properties 
in the electron-hole asymmetric case.

\begin{figure}[tb]
\begin{center}
\leavevmode
\includegraphics[width=0.9\linewidth]{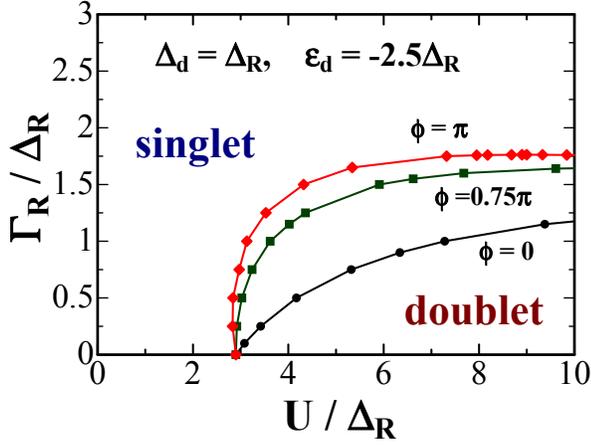}
\caption{
Phase diagrams of the ground state 
away from half-filling for several values of the phase difference $\phi$, 
where  $\epsilon_d = -2.5|\Delta_R|$ and $\Delta_d =|\Delta_R|$.
}
\label{fig:Phase_diagram_asymetric}
\end{center}
\end{figure}

\begin{figure}[tb]
\begin{center}
\leavevmode
\includegraphics[width=0.9\linewidth]{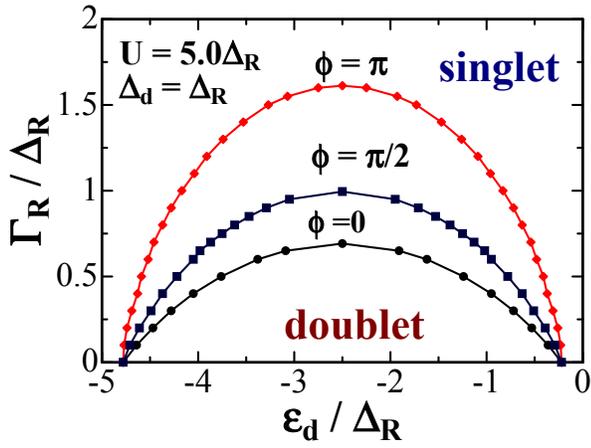}
\caption{
Phase diagrams of the ground state 
as a function of $\epsilon_d$ and $\Gamma_R$,
where $\Delta_d =|\Delta_R|$
 and $U=5.0|\Delta_R|$. 
}
\label{fig:Phase_diagram_asymetric_largeU}
\end{center}
\end{figure}

\begin{figure}[tb]
\begin{center}
\leavevmode
\includegraphics[width=0.9\linewidth]{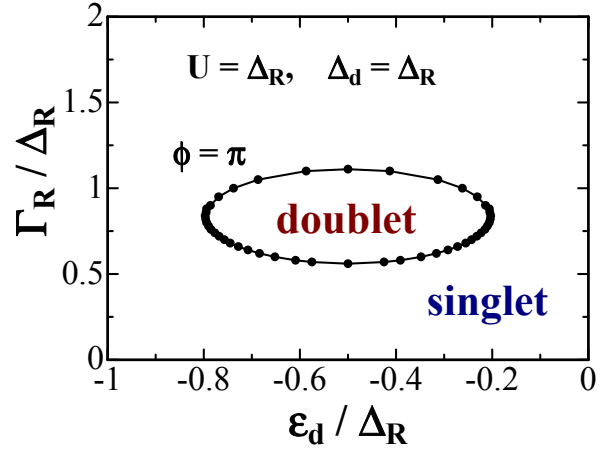}
\caption{
Phase diagrams of the ground state
as a function of $\epsilon_d$ and $\Gamma_R$,
where $\phi=\pi$, $\Delta_d =|\Delta_R|$
 and $U=1.0|\Delta_R|$. 
Note that the scale of the horizontal axis is different  
from that of Fig.\ \ref{fig:Phase_diagram_asymetric_largeU}.
}
\label{fig:Phase_diagram_asymetric_smallU}
\end{center}
\end{figure}

In Fig.\ \ref{fig:Phase_diagram_asymetric}, 
one example of the phase diagram 
is plotted as a function  of $U$ and $\Gamma_R$, 
where $\epsilon_d = -2.5|\Delta_R|$ 
and  $\Delta_d =|\Delta_R|$. 
The model has the electron-hole symmetry at $U=5.0|\Delta_R|$ 
and the results coincide with that 
at the same $U$ in Fig.\ \ref{fig:Phase_diagram_josephson}(a). 
As expected, the region where the doublet state stabilized 
becomes narrow away from half-filling. 
Particularly the re-entrant region of the doublet state, 
which is seen at half-filling, almost vanishes 
in the case of Fig.\ \ref{fig:Phase_diagram_asymetric}. 
At $\Gamma_R=0$, all the lines for different $\phi$ 
start from $U_{\text{a}} = (\epsilon_d^2 + \Delta_d^2)/(-\epsilon_d)$,
where the level crossing occurs in the atomic limit. 
Then, for large $U$, the boundaries become flat.
We next consider the $\epsilon_d$ dependence of the 
ground state.
In the following two figures,
 the phase diagrams are plotted in a $\Gamma_R$ vs $\epsilon_d$ plane
for two different values of $U$: 
it is chosen to be  $U=5.0|\Delta_R|$ and $U=1.0|\Delta_R|$ in
Figs.\ \ref{fig:Phase_diagram_asymetric_largeU} 
and \ref{fig:Phase_diagram_asymetric_smallU}, respectively. 
Here the impurity SC gaps is taken to be $\Delta_d =|\Delta_R|$.
The figure \ref{fig:Phase_diagram_asymetric_largeU} 
is an example of the phase diagram for large $U$, where
the impurity is half-filled  at $\epsilon_d = -2.5|\Delta_R|$,
and the doublet phase vanishes when $\epsilon_d$ or $\epsilon_d+U$ 
approaches the Fermi level at $\mu=0$.
The doublet region becomes wider with increasing $\phi$,
and this tendency is maximized at $\phi=\pi$.
For small $U$,   
the re-entrant doublet region appears 
as an {\em island} in the phase diagram  
 Fig.\ \ref{fig:Phase_diagram_asymetric_smallU}, 
where the line for $\phi=\pi$ is shown. 
The magnetic doublet state is most stable 
at half-filling  $\epsilon_d = -0.5|\Delta_R|$.
In this re-entrant region, the bound state lies close to the Fermi level.

\begin{figure}[tb]
\begin{center}
\leavevmode
\includegraphics[width=0.9\linewidth]{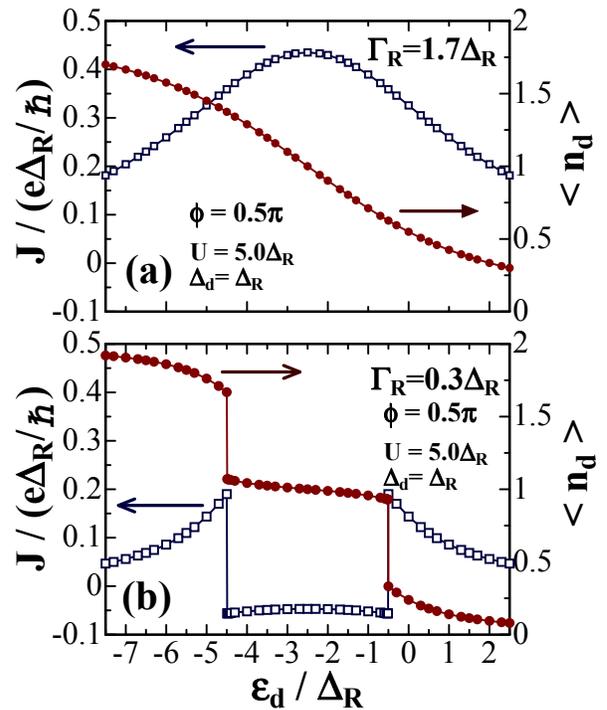}
\caption{
The Josephson current and the number of the electrons in the impurity  
$\langle n_d \rangle$ as 
a function of $\epsilon_d$ for 
(a) $\Gamma_R = 1.7|\Delta_R|$ and
(b) $\Gamma_R = 0.3|\Delta_R|$. 
The parameters are taken to be $\phi= 0.5 \pi$, 
$\Delta_d =|\Delta_R|$, and $U=5.0|\Delta_R|$.
The ground state is a singlet in the case (a),
while in (b) the quantum transition occurs 
and the doublet state is stabilized 
around half-filling. 
}
\label{fig:J_nd_asm}
\end{center}
\end{figure}

We have also calculated the Josephson current and 
the average number of electrons 
in the impurity site $\langle n_d \rangle$.
In Fig.\ \ref{fig:J_nd_asm}, the results 
are plotted as a function of $\epsilon_d$ for 
(a) $\Gamma_R = 1.7|\Delta_R|$ and 
(b) $\Gamma_R = 0.3|\Delta_R|$.
The parameters are taken to be $\phi= 0.5 \pi$, 
$\Delta_d =|\Delta_R|$, and $U=5.0|\Delta_R|$: 
the ground state for these parameters are shown
in Fig.\ \ref{fig:Phase_diagram_asymetric_largeU}. 
In the case of (a), the ground state is a singlet independent of $\epsilon_d$. 
The current shows a maximum at half-filling 
 $\epsilon_d = -2.5|\Delta_R|$, 
and $\langle n_d \rangle$ decreases monotonically 
with increasing $\epsilon_d$.
On the other hand, in the case of (b),
the quantum phase transition occurs near half-filling, 
where the doublet state is stabilized.
The expectation values show the singular behavior 
at the critical point, and particularly 
the Josephson current becomes small and changes the direction.

\section{Summary}
\label{sec:summary}

In summary, 
we have studied the ground-state properties of a model 
for the quantum dot embedded in the Josephson junction 
making use of the NRG method. To carry out precise calculations 
in a wide parameter range, we have introduced a 
simplified single-channel model that has 
an extra superconducting gap $\Delta_d$ at the impurity site. 
This model captures the essence of 
the Josephson and Kondo physics, and can be derived 
from the two-channel model when one of the bulk SC gaps is
 much larger than the others $|\Delta_L| \gg |\Delta_R|$. 
This kind of asymmetry in the Josephson couplings 
causes the nontrivial changes in the superconducting
proximity effects on the impurity site, and it affects 
the low-energy spin dynamics of the quantum impurity.

The numerical results for the simplified model 
show that the Josephson phase 
tends to disturb the conduction electron screening of  
the local moment, and destabilizes the paramagnetic 
singlet ground state to drive a quantum phase transition to 
the magnetic doublet ground state. 
At the transition point, the direction of 
the Josephson current changes discontinuously due to 
the spin-flip tunneling.
Specifically, when the Josephson phase is $\phi \simeq\pi$ and 
the other parameters are set so that the 
conductance in a normal limit $\Delta_{R},\, \Delta_{L} \to 0$
is close to the unitary limit value,
the bound state comes near the Fermi level. 
This zero mode causes a re-entrant magnetic phase transition 
for the asymmetric Josephson couplings.

  It is known for the Anderson impurity in normal metals 
  that the magnetic transition described by the mean-field theory 
  is an artifact due to the approximation, 
  i.e., quantum fluctuations stabilize 
  the singlet ground state. 
  In the superconducting case, however, the gap changes 
  the structures of the low-energy excitations, and makes 
  the occurrence of the quantum phase transition possible,
  the phase diagrams for which are provided in the present paper.
  The asymmetries in the Josephson couplings and 
  the phase difference of the gaps add interesting varieties 
  in the ground-state properties of the Kondo system 
  through the changes in the bound state.

\section*{Acknowledgements}
We are  grateful to O.\ Sakai and 
Y.\ Shimizu for valuable discussions. 
One of us (AO) wishes to acknowledge 
a support by the Grant-in-Aid 
for Scientific Research from JSPS.
ACH wishes to thank the EPSRC(Grant GR/S18571/01) 
for financial support.
Numerical computation was partly performed 
at computation center of Nagoya University and 
at Yukawa Institute Computer Facility.

\appendix

 \section{Functional method}
 \label{sec:action}

 The impurity part of the partition function 
 can be expressed using the path integral formulation as
 \begin{align}
 & Z_{\text{eff}} \,=\,   
 \int \!
 {\cal D}\eta^{\dagger} {\cal D}\eta \  \text{e}^{- S_{\text{eff}} }
 \;, 
 \label{eq:Z}
 \\
   & S_{\text{eff}}  \,=   
    \int_{0}^{\beta} \!\! \int_{0}^{\beta} 
    \text{d}\tau \, \text{d}\tau'  \!\!\!
 \begin{array}{l}
   \\
 \left[
  \begin{matrix}
  \eta_{\uparrow}^{\dagger}(\tau) \cr
  \eta_{\downarrow}^{\phantom{\dagger}}(\tau) \cr
 \end{matrix}
  \right] 
 \end{array}
\!\!\!\! \mbox{\boldmath $K$}_{dd}^0(\tau, \tau') 
  \left[
    \eta_{\uparrow}^{\phantom{\dagger}}(\tau') \   
  \eta_{\downarrow}^{\dagger}(\tau')
 \right] 
 \nonumber \\
 & \qquad \  -\,\frac{U}{2} \! \int_{0}^{\beta} \! \text{d}\tau \,
  \left(\, 
  \sum_{\sigma} 
  \eta_{\sigma}^{\dagger}(\tau)\,
  \eta_{\sigma}^{\phantom{\dagger}}(\tau)\,
  \,-\,1\,\right)^2
  \,, 
   \label{eq:action_tot}
 \\
  &   \mbox{\boldmath $K$}_{dd}^0(\tau,\tau') \ = \ 
 \frac{1}{\beta} \sum_{\omega_n}
 \left\{ \mbox{\boldmath $G$}_{dd}^0(\text{i} \omega_n) \right\}^{-1} 
     \, \text{e}^{- \text{i} \omega_n (\tau-\tau')}
  \;, 
   \label{eq:action_0}
  \end{align}
where $\eta_{\sigma}^{\phantom{\dagger}}(\tau)$ is 
a Grassmann number. 
The information about 
the conduction electrons enters into $S_{\text{eff}}$ 
through the noninteracting Green's 
function $\mbox{\boldmath $G$}_{dd}^0(\text{i} \omega_n)$.
Thus, if the different Hamiltonians give 
the same $\mbox{\boldmath $G$}_{dd}^0(\text{i} \omega_n)$,
they also describe the same local properties 
around the impurity.\cite{AO_highbias}

\section{Bogoliubov-de Gennes representation}
\label{sec:bogoliubov}

We provide here the expression of the Green's function 
in terms of the Bogoliubov quasiparticles, 
which will be used in Appendix \ref{sec:degenerate_perturbation}
to carry out the perturbative calculations.  
The noninteracting Hamiltonian, 
$
 {\cal H}^0  
 \equiv      {\cal H}_{d}^0 
  +  \sum_{\lambda=L,R} \left(\, {\cal H}_{\lambda} 
  +  {\cal H}_{d\lambda}^{T} \,\right)
 $,
can be diagonalized  
using the Bogoliubov operator $\gamma_{n\sigma}$;
\begin{equation}
{\cal H}_0 \,=\, 
E_b 
\sum_{\sigma} 
\gamma_{0\sigma}^{\dagger} \gamma_{0\sigma}^{\phantom{\dagger}}
\,+\,
\sum_{\sigma} 
\sum_{n > 0} 
E_n 
\gamma_{n\sigma}^{\dagger} \gamma_{n\sigma}^{\phantom{\dagger}}
\ + \ \mbox{const.} 
\;.
\label{eq:BdJ}
\end{equation}
Here the label $n=0$ is assigned to the bound state in the gap, 
and that for $n>0$ corresponds to the 
excitations in the continuum $E_n \geq \min(|\Delta_L|,\,|\Delta_R|)$.
Then the impurity Green's function can be written as,
\begin{align}
\mbox{\boldmath $G$}_{dd}^0(\text{i} \omega_l)
\,& =\,
 \mbox{\boldmath $U$}_{d0} \,
{\text{i} \omega_l \mbox{\boldmath $1$} \,+\,
 E_b  \mbox{\boldmath $\tau$}_3 
 \over 
 -\,\omega_l^2 \,-\,E_b^2 }\  
 \mbox{\boldmath $U$}_{d0}^{\dagger} 
\nonumber 
\\ 
& \ \, +\, 
\sum_{n>0}\, 
 \mbox{\boldmath $U$}_{dn} \,
{\text{i} \omega_l \mbox{\boldmath $1$} \,+\,
 E_n \mbox{\boldmath $\tau$}_3 
 \over 
 -\,\omega_l^2 \,-\,E_n^2 }\  
 \mbox{\boldmath $U$}_{dn}^{\dagger} 
  \;.
  \label{eq:G_with_uv}
 \end{align}
Here, $\mbox{\boldmath $U$}_{dn}$ is the matrix   
defined by the Unitary transformation
\begin{align}
\begin{array}{l}
\\
 \left[
 \begin{matrix}
  d_{\uparrow}^{\phantom{\dagger}} \cr
  d_{\downarrow}^{\dagger}  \cr
 \end{matrix}
 \right] 
\end{array} 
&\,= \, 
\mbox{\boldmath $U$}_{d0}
\!\!
\begin{array}{l}
     \\
 \left[
 \begin{matrix}
  \gamma_{0\uparrow}^{\phantom{\dagger}} \cr
  \gamma_{0\downarrow}^{\dagger} \cr
 \end{matrix}
 \right] 
\end{array} 
\ + \ 
\sum_{n>0}
\,\mbox{\boldmath $U$}_{dn}
\!\!
\begin{array}{l}
     \\
 \left[
 \begin{matrix}
  \gamma_{n\uparrow}^{\phantom{\dagger}} \cr
  \gamma_{n\downarrow}^{\dagger}  \cr
 \end{matrix}
 \right] 
\end{array}  
\;,
\label{eq:d_vs_bogoron}
\\
\mbox{\boldmath $U$}_{dn} &\, = \,
\left[ \,
\begin{matrix} 
u_{dn}  &  -v_{dn}^* \cr
 v_{dn} & \phantom{-}u_{dn}^* \cr
\end{matrix} 
\,\right]  \;,
\end{align}
and it has a property
$
\mbox{\boldmath $U$}_{dn}\mbox{\boldmath $U$}_{dn}^{\dagger} 
= \bigl(\, \left| u_{dn} \right|^2 +  
\left| v_{dn} \right|^2 \,\bigr) \mbox{\boldmath $1$}
$.
With these quasiparticle states,
the SC correlation for noninteracting electrons 
$\chi_d^0 = \langle d_{\downarrow} d_{\uparrow} \rangle^0$ 
can be expressed as 
\begin{align}
\chi_d^0
 \,& = \,   
-\,
u_{d0} v_{d0}^* 
 \left\{ \,
1\, - \, 2f(E_b) 
\,\right\} 
\nonumber
\\ 
 & \ \ \  - \, 
\sum_{n>0}\, 
\,u_{dn} v_{dn}^* 
\left\{\,
1\, - \, 2f(E_n) 
\,\right\}  \;.
 \label{eq:chi_Lehmann}
\end{align}
Note that the first term in the right-hand side, 
which corresponds to the contributions of the bound state,
vanishes if the bound state appears at the Fermi energy $E_b=0$.

\section{Correlation functions for the zero mode $E_b=0$ 
}
\label{sec:pair_potential_at_E}

As mentioned in \S \ref{sec:bound_state},
the bound state appears just on the Fermi level 
when $\phi=\pi$, $\Gamma_R=\Gamma_L$  ($\equiv \Gamma$), 
and $E_d=0$.
Around the pole at $E_b=0$,  
the Green's function can be written in the form
\begin{align}
\mbox{\boldmath $G$}_{dd}^0(\epsilon)
&\,\simeq\,   
 \frac{a}{\epsilon + \text{i} 0^+} \,\mbox{\boldmath $1$}\;, \\
a &\,=\, 
{1 \over 1 + \frac{\Gamma}{|\Delta_L|} + \frac{\Gamma}{|\Delta_R|} }
\;.
\label{eq:pole1}
\end{align}
The spectral weight at the impurity site, $a$, decreases with 
increasing $\Gamma$ because the 
bound-state wavefunction penetrates deep inside the leads 
for large $\Gamma$.
Alternatively, the Green's function can be expressed 
 using  Eq.\ (\ref{eq:G_with_uv}) as 
\begin{equation}
\mbox{\boldmath $G$}_{dd}^0(\epsilon)
 \,\simeq \,
 { 
 |u_{d0}|^2+|v_{d0}|^2
 \over
 \epsilon + \text{i} 0^+  
 } \,
  \mbox{\boldmath $1$} 
\; .   
\label{eq:pole2}
\end{equation}
Therefore comparing Eqs.\ (\ref{eq:pole1}) and (\ref{eq:pole2}), we have
$ a =  |u_{d0}|^2+|v_{d0}|^2$. 
Furthermore, it is concluded that
$|u_{d0}|^2=|v_{d0}|^2 = a/2$  because of the electron-hole symmetry.
Also, $u_{d0}$ and $v_{d0}$ are essentially real at $\phi = \pi$.

In the presence of the zero mode the expression of
the SC correlation, Eq.\ (\ref{eq:chi0}), can be 
rewritten in the form  
\begin{align}
 & \chi_{d}^0 \ = 
\nonumber 
\\ 
& \ \ \ \    
 \frac{1}{\beta} 
 \sum_{\omega_n}
{ \Gamma \, \omega_n^2 \over \det 
\left\{ \mbox{\boldmath $G$}_{dd}^{0}(\text{i}\omega_n) \right\}^{-1} 
 }
\, 
{  
|\Delta_L|^2 \, -\,
|\Delta_R|^2
\over 
\left( \omega_n^2 + |\Delta_{L}|^2 \right)
\left(\omega_n^2 + |\Delta_{R}|^2 \right)
} 
\nonumber \\
& 
\ \ \ \ 
\times 
\left[\,
{  
|\Delta_L|
\over 
\sqrt{\omega_n^2 + |\Delta_{L}|^2 }
} 
\ + \  
{
|\Delta_R|
\over 
\sqrt{\omega_n^2 + |\Delta_{R}|^2 }
} 
\,\right]^{-1}  \;.
\label{eq:chi0_pi}
\end{align}
This expression clearly shows that
$\chi_d^0$ vanishes for $|\Delta_{L}|=|\Delta_{R}|$.  
Note that
the behavior of the determinant 
 near the bound state is deduced form Eq.\ (\ref{eq:pole1}) 
as
$
  \det \left\{ \mbox{\boldmath $G$}_{dd}^{0}(\text{i}\omega_n) \right\}^{-1} 
\simeq  -\omega_n^2/a^2$. 
 In  Eq.\ (\ref{eq:chi0_pi}) this $\omega_n^2$ dependence is 
 canceled out by that in the numerator.

\section{Perturbation theory for 
the zero mode
}
\label{sec:degenerate_perturbation}

In the presence of the zero mode, i.e.,  
 for $\phi=\pi$, 
$\Gamma_L =\Gamma_R$ ($\equiv \Gamma$),
and $E_d=0$, the following four states are degenerate; 
\begin{align}
|\text{I} \rangle &\,=\, |\widetilde{0} \rangle  \;, \\
|\text{II} \rangle &\,=\, \gamma_{0\uparrow}^{\dagger}|\widetilde{0} \rangle  
\;, \\
|\text{III} \rangle &\,=\, 
\gamma_{0\downarrow}^{\dagger}|\widetilde{0} \rangle 
 \;, \\
|\text{IV} \rangle & \,=\, 
\gamma_{0\uparrow}^{\dagger}\gamma_{0\downarrow}^{\dagger}
|\widetilde{0} \rangle  \;. 
\end{align}
Here  $|\widetilde{0} \rangle$ is the vacuum state with respect 
to the Bogoliubov 
particles $\gamma_{n \sigma} |\widetilde{0} \rangle  =  0$.
An infinitesimal repulsion  ${\cal H}_{d}^{U} = 
(U/2)\, \left(  n_{d} -1 \right)^2$ lifts the degeneracy of the ground state.
We have calculated the matrix elements  
$\langle \alpha | {\cal H}_{d}^{U}  |\alpha' \rangle$
in this subspace  
making use of Eq.\ (\ref{eq:d_vs_bogoron}) 
and the equations provided in Appendix \ref{sec:pair_potential_at_E}.
We found that the off-diagonal elements are zero
\begin{equation}
 \langle \alpha | {\cal H}_{d}^{U}  |\alpha' \rangle
 \,=\, 0 \;, \ \ \mbox{for} \ \ \alpha \neq \alpha' \;,
\end{equation}
and the energy shift is determined simply by the diagonal elements 
\begin{align}
 \langle \text{I} | {\cal H}_{d}^{U}  |\text{I} \rangle
-
 \langle \text{II} | {\cal H}_{d}^{U}  |\text{II} \rangle
& \, = 
\frac{U}{2} 
\left[ a^2 -  2a \chi_d^0\,  \mbox{sgn}\,(u_{d0} v_{d0})
\right] , 
\label{eq:singleI_double}\\
 \langle \text{IV} | {\cal H}_{d}^{U}  |\text{IV} \rangle
-
 \langle \text{II} | {\cal H}_{d}^{U}  |\text{II} \rangle
& \, =  
\frac{U}{2} 
\left[a^2 +  2a \chi_d^0\,  \mbox{sgn}\,(u_{d0} v_{d0})
\right] .
\label{eq:singleIV_double}
\end{align}
Note that 
$E_{S=1/2} \equiv  
\langle \text{II} | {\cal H}_{d}^{U}  |\text{II} \rangle
  =   \langle \text{III} | {\cal H}_{d}^{U}  |\text{III} \rangle$ 
is the energy shift for the doublet states.
Consequently, the energy difference between 
the doublet and the lowest singlet states is given by
\begin{equation}
E_{S=0}
- E_{S=1/2}
\,=\,
\frac{U}{2} 
\,\left[\,a^2\, \,-\, 2\,a\left|\chi_d^0\right|\, 
\,\right] 
\;.
\end{equation}
Thus the level crossing occurs at $a - 2|\chi_d^0|=0$, where
$\chi_d^0$ is given  by  Eq.\ (\ref{eq:chi0_pi}), and 
this equation determines the phase boundary 
between the doublet and singlet states. 
The results are shown in Fig.\ \ref{fig:gam_critical}
as a function of the ratio $x \equiv |\Delta_R|/|\Delta_L|$. 
At $x=0$, i.e., in the limit of $|\Delta_L| \to \infty$,
 Eq.\ (\ref{eq:chi0_pi}) simplifies for $T=0$ as
\begin{align}
& \chi_d^0 \ =  
\, -\int_{-\infty}^{\infty}
\frac{\text{d}\omega}{2\pi}\ \Gamma 
\nonumber
\\
& \qquad \quad \times
\biggl[\, 
\left( \sqrt{\omega^2 + |\Delta_R|^2 } +2\Gamma \right)
\left( \sqrt{\omega^2 + |\Delta_R|^2 } + |\Delta_R| \right)
\nonumber 
\\
& \qquad \qquad \quad 
+ \  2\, \Gamma^2 \,
\biggr]^{-1}
\;,
\end{align}
and the critical value of $\Gamma$ is given by 
$\Gamma_{\text{cr}}/|\Delta_R| = 1.77064 \cdots$.  
In the opposite limit at $x=1-0^+$, i.e., 
for $|\Delta_L|=|\Delta_R|$, the critical value
diverges exponentially as 
$ {\Gamma_{\text{cr}}}/{|\Delta_R|}
 \simeq
 4\,\text{e}^{[\, \pi/(1-x) \,+ \,1 \,]}\,$,
 so that the doublet ground state is stabilized for any values of $\Gamma$.


\begin{thebibliography}{99}




\bibitem{SodaMatsuuraNagaoka}
T. Soda, T. Matsuura, and Y. Nagaoka:
Prog.\ Theor.\ Phys.\ {\bf 38} (1967) 551.

\bibitem{Matsuura}
T. Matsuura:
Prog.\ Theor.\ Phys.\ {\bf 57} (1977) 1823.


\bibitem{MullerHartmanZittartz}
M\"{u}ller-Hartman and J. Zittartz:
Z.\ Phys.\ {\bf 234} (1970) 58.



\bibitem{Hewson}
A. C. Hewson: 
{\em  The Kondo Problem to Heavy Fermions\/} 
(Cambridge University Press, Cambridge, 1993). 




\bibitem{PWA}
P. W. Anderson:
Phys.\ Rev.\  \textbf{124} (1961) 41. 








\bibitem{JarrelSiviaPatton}
M. Jarrell, D. S. Sivia and B Patton, 
Phys.\ Rev.\ B \textbf{42} (1990) 4804. 




\bibitem{SSSS}
K. Satori, H. Shiba, O. Sakai and 
Y. Shimizu:  
J.\ Phys.\ Soc.\ Jpn.\ \textbf{61} (1992) 3239. 

\bibitem{SSSS_2}
O. Sakai, Y. Shimizu, H. Shiba, and K. Satori:
J.\ Phys.\ Soc.\ Jpn.\ \textbf{62} (1993) 3181. 



\bibitem{YoshiokaOhashi}
T. Yoshioka and Y. Ohashi:  
J.\ Phys.\ Soc.\ Jpn.\ \textbf{69} (2000) 1812.


\bibitem{MatsumotoKoga}
M. Matsumoto and M. Koga:  
J.\ Phys.\ Soc.\ Jpn.\ \textbf{70} (2001) 2860.


\bibitem{GlazmanMatveev}
L. I. Glazman and K. A. Matveev:
Pis'ma Zh.\  Eksp.\ Teor.\ Fiz.\ \textbf{49} (1989) 570
[JETP Lett.\ \textbf{49} (1989) 659].



\bibitem{Beenakker}
C. W. J. Beenakker: 
in {\em Transport Phenomena in Mesoscopic Systems},
 edited by H. Fukuyama and T. Ando (Springer-Verlag, Berlin, 1992).




\bibitem{Kulik}
I. O. Kulik:
Soviet Physics JETP \textbf{22} (1966) 841.


\bibitem{ShibaSoda}
H. Shiba and T. Soda:  
Prog.\ Theor.\ Phys. \textbf{41} (1969) 25.



 
\bibitem{SpivakKivelson}
B. I. Spivak and S. A. Kivelson: 
Phys.\ Rev.\ B \textbf{43} (1991) 3740.



\bibitem{Ishizaka}
S. Ishizaka, J. Sone and T. Ando:  
Phys.\ Rev.\ B \textbf{52} (1995) 8358.



\bibitem{ClerkAmbegaokar}
A. A. Clerk and V. Ambegaokar:  
Phys.\ Rev.\ B \textbf{61} (2000) 9109.


\bibitem{RozhkovArovas}
A. V. Rozhkov and D. P. Arovas: 
Phys.\ Rev.\ Lett.\ \textbf{82} (1999) 2788.


\bibitem{Avishiai}
Y. Avishai, A. Golub and A. D. Zaikin:  
Phys.\ Rev.\ B \textbf{67} (2003) 041301.

\bibitem{Vecino}
E. Vecino, A. Mart\'{i}n-Rodero, and A. Levy Yeyati:
Phys.\ Rev.\ B \textbf{68} (2003) 035105.


\bibitem{Kusakabe}
K. Kusakabe, Y. Tanaka and Y. Tanuma:
Physica E, \textbf{18} (2003) 50.



\bibitem{Choi}
M-S. Choi, M. Lee, K. Kang, and W. Belzig:  cond-mat/031227.


\bibitem{KWW}
H. R. Krishna-murth, J. W. Wilkins: 
and K. G. Wilson, Phys.\ Rev.\ B {\bf 21} (1980) 1003. 



\bibitem{SakaiShimizuKasuya}
O. Sakai, Y. Shimizu, and T. Kasuya:
Prog.\ Theor.\ Phys.\ Suppl.\ {\bf 108} (1992) 73.



\bibitem{Hewson_renorm}
A.\ C.\ Hewson: J.\ Phys.:\ Condens.\ Matter {\bf 13} (2001) 10011.


\bibitem{AO_highbias}
A. Oguri,  
J.\ Phys.\ Soc.\ Jpn.\ \textbf{71} (2002) 2696.

\end{thebibliography}
\end{document}